\documentclass[11pt]{article}
\pdfoutput=1
\usepackage{jcappub} 

\usepackage{lineno}
\usepackage{subcaption}
\usepackage{xcolor}

\definecolor{color1}{rgb}{0.51, 0, 0.88}

\hypersetup{linkcolor=color1,citecolor=color1,filecolor=color1,urlcolor=color1}

\usepackage{booktabs}
\usepackage{adjustbox}

\newcommand{\jcap}{J.\ Cosmol.\ Astropart.\ Phys.}
\newcommand{\aap}{Astron.\ Astrophys.}
\newcommand{\prd}{Phys.\ Rev.\ D}
\newcommand{\prl}{Phys.\ Rev.\ Lett.}
\newcommand{\apj}{Astrophys.\ J.}
\newcommand{\apjs}{Astrophys.\ J.\ Suppl.}
\newcommand{\apjl}{Astrophys.\ J.\ Lett.}
\newcommand{\mnras}{Mon.\ Not.\ R.\ Astron.\ Soc.}

\title{Constraining primordial oscillations and inflationary particle production with $\boldsymbol{Planck}$, ACT DR6, and DESI DR2}

\usepackage{orcidlink}

\author[1, 2]{Simran K. Nerval\orcidlink{0009-0006-0076-2613},}

\author[1, 2]{Ren\'{e}e Hlo\v{z}ek\orcidlink{0000-0002-0965-7864},}

\author[3]{Hidde T. Jense\orcidlink{0000-0002-9429-0015}}

\author[4]{and J.~Richard~Bond\orcidlink{0000-0003-2358-9949}}

\affiliation[1]{David A. Dunlap Department of Astronomy \& Astrophysics, University of Toronto, 50 St. George St., Toronto, ON M5S 3H4, Canada} 
\affiliation[2]{Dunlap Institute for Astronomy and Astrophysics, University of Toronto, 50 St. George St., Toronto, ON M5S 3H4, Canada}

\affiliation[3]{School of Physics and Astronomy, Cardiff University, Queen’s Buildings, The Parade, Cardiff CF24 3AA, United Kingdom}

\affiliation[4]{Canadian Institute for Theoretical Astrophysics, University of Toronto, 60 St. George St., Toronto, ON M5S 3H8, Canada}

\emailAdd{simran.nerval@mail.utoronto.ca}

\abstract{Non-standard inflationary models often predict oscillatory features in the primordial power spectrum. We present constraints on general oscillatory templates for primordial power spectra, including those that vary linearly and logarithmically with wavenumber, as well as oscillations induced by inflationary particle production. We utilize the \textit{Planck} 2018 and Atacama Cosmology Telescope Data Release 6 cosmic microwave background data as well as large-scale structure data from the Dark Energy Spectroscopic Instrument Data Release 2. To efficiently explore the multimodal posteriors for these models as well as performing mode comparisons, we integrate the preconditioned sequential Monte Carlo sampler, \texttt{pocoMC}, into the widely used sampling code, \texttt{Cobaya}. We find that the combined dataset tightens the 95\% CL upper bounds on the general oscillation amplitudes to $A_{\text{lin}} < 0.021$, $A_{\text{log}} < 0.022$, and $A_{\text{log rf}} < 0.023$, restricting the amplitude to $\sim 2\%\, A_s$. For the inflationary particle production model, our analysis places a maximum a posteriori constraint on the coupling constant of $g = 0.034$. While these models all provide an improved fit to the data compared to the concordance $\Lambda$CDM model, the Bayesian evidence still reveals a moderate preference for $\Lambda$CDM compared to models with general oscillations and is inconclusive regarding the particle production model, suggesting that the added complexity of these models beyond the standard model is not statistically justified by current data.}


\begin{document}
\maketitle
\flushbottom

\section{Introduction}
\label{sec:intro}

The cosmic microwave background (CMB) provides a window into the earliest moments of the universe. In our standard cosmological picture, we have an epoch of exponential expansion known as inflation, which explains the homogeneity and isotropy of our universe \citep{1978AnPhy.115...78B, 1980ApJ...241L..59K, Starobinsky:1980te, Guth, 1982PhLB..117..175S, 1982PhLB..108..389L, 1982PhRvL..48.1220A}. Additionally, it seeded the initial density perturbations, which eventually formed the large-scale structure we observe today. Standard single-field slow-roll inflationary models predict a nearly scale-invariant and featureless primordial power spectrum (PPS), which we describe in terms of a power law \citep{Pk1995PhRvD..52.1739K}:
\begin{equation}
    P_{\mathcal{R}}(k) = A_s\bigg(\frac{k}{k_*}\bigg)^{n_s - 1 \ + \ (1/2)\,(dn_s/d\ln k) \, \ln(k/k_*)},
\end{equation}
\noindent where $A_s$ is the amplitude of the primordial scalar perturbations, $n_s$ is the scalar spectral index, $k_* = 0.05 \: \text{Mpc}^{-1}$ is the pivot scale, and $dn_s/d\ln k$ is the running of the scalar spectral index with wavenumber. This standard power law model is broadly consistent with precise CMB measurements from telescopes such as the \textit{Planck} satellite \citep{planckoverview2020A&A...641A...1P, plancklikelihood2020A&A...641A...5P}, the Atacama Cosmology Telescope (ACT) \citep{swetz/etal:2011,thornton/etal:2016,henderson/etal:2016, ACTlcdm2025JCAP...11..062L}, and the South Pole Telescope \citep{SPT2026PhRvD.113h3504C}.
\begin{figure}[htbp!]
	\centering
\includegraphics[width=0.6\textwidth]{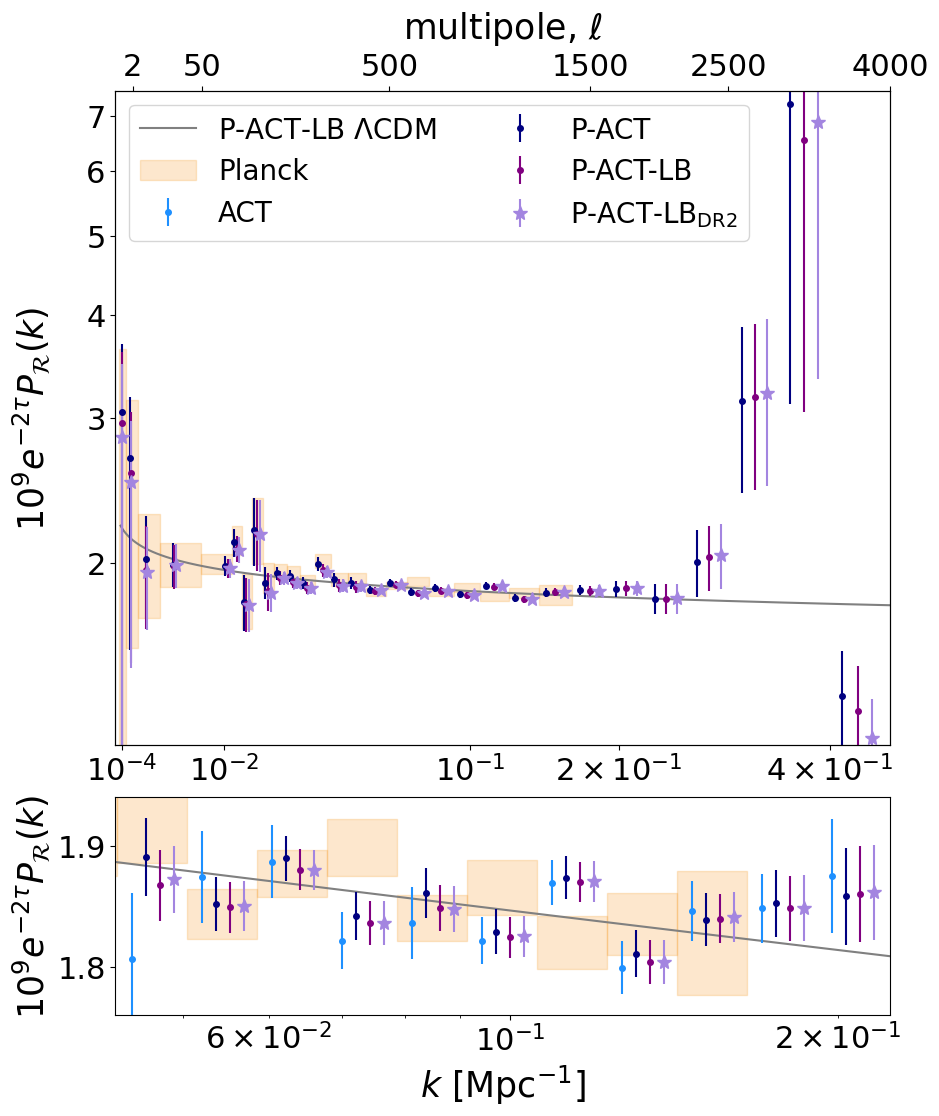}
	\caption{Binned reconstruction of the dimensionless primordial curvature power spectrum with 68\% CL errors for different datasets, updated from \cite{ACTextended2025JCAP...11..063C}. The ACT (P-ACT) wavenumbers on the $x$-axis are shifted to the left 4\% (1\%) and the P-ACT-LB (P-ACT-LB$_{\text{DR2}}$) to the right by 2\% (5\%) to expose the central values. The top panel shows the reconstruction over the full range of wavenumbers considered, highlighting the improvement achieved with the addition of the ACT data (navy versus orange bands). For this panel, the $x$-axis is scaled as $k^{0.5}$ in order to best show the small scales. The bottom panel zooms into the region where ACT alone (light blue) becomes comparable and then overtakes \textit{Planck} in constraining power. Adding lensing and BAO data (purple) has minimal impact. The updated constraints using DESI DR2 (P-ACT-LB$_{\text{DR2}}$) are shown as stars (light purple). The binned measurement is consistent with the P-ACT-LB best-fit $\Lambda$CDM power-law spectrum shown in grey.}
	\label{fig:binned Pk}
\end{figure}
Despite these precise cosmological measurements, the physics driving inflation remains poorly constrained. Open questions include which field(s) drove inflation, what the shape of the potential is, and what possible departures there are in the model from single-field slow-roll dynamics. Our current upper bound on the energy scale of inflation comes from the \textit{Planck} PR4 and BICEP/Keck upper bound of the tensor-to-scalar ratio, $r \equiv A_t/A_s < 0.032$, which corresponds to $E_{\text{inf}} < 1.3 \times 10^{16}$ GeV at 95\% confidence \citep{Tristram2022PhRvD.105h3524T}. At these energy scales, the Standard Model of particle physics is likely incomplete, and additional fields or interactions are expected to become relevant. Many of these ``exotic" inflationary models, such as axion monodromy inflation \citep{Flauger:2014ana, 2017JCAP...10..058F}, unwinding inflation \citep{unwinding2013JCAP...03..004D}, trans-Planckian effects \cite{transplanck2003PhRvD..68f3513M}, and inflationary particle production \citep{Barnaby2010, 1702.07661.2017JCAP...05..054P}, predict oscillatory features in the PPS. 

\begin{figure}
\centering
\includegraphics[width=0.9\linewidth]{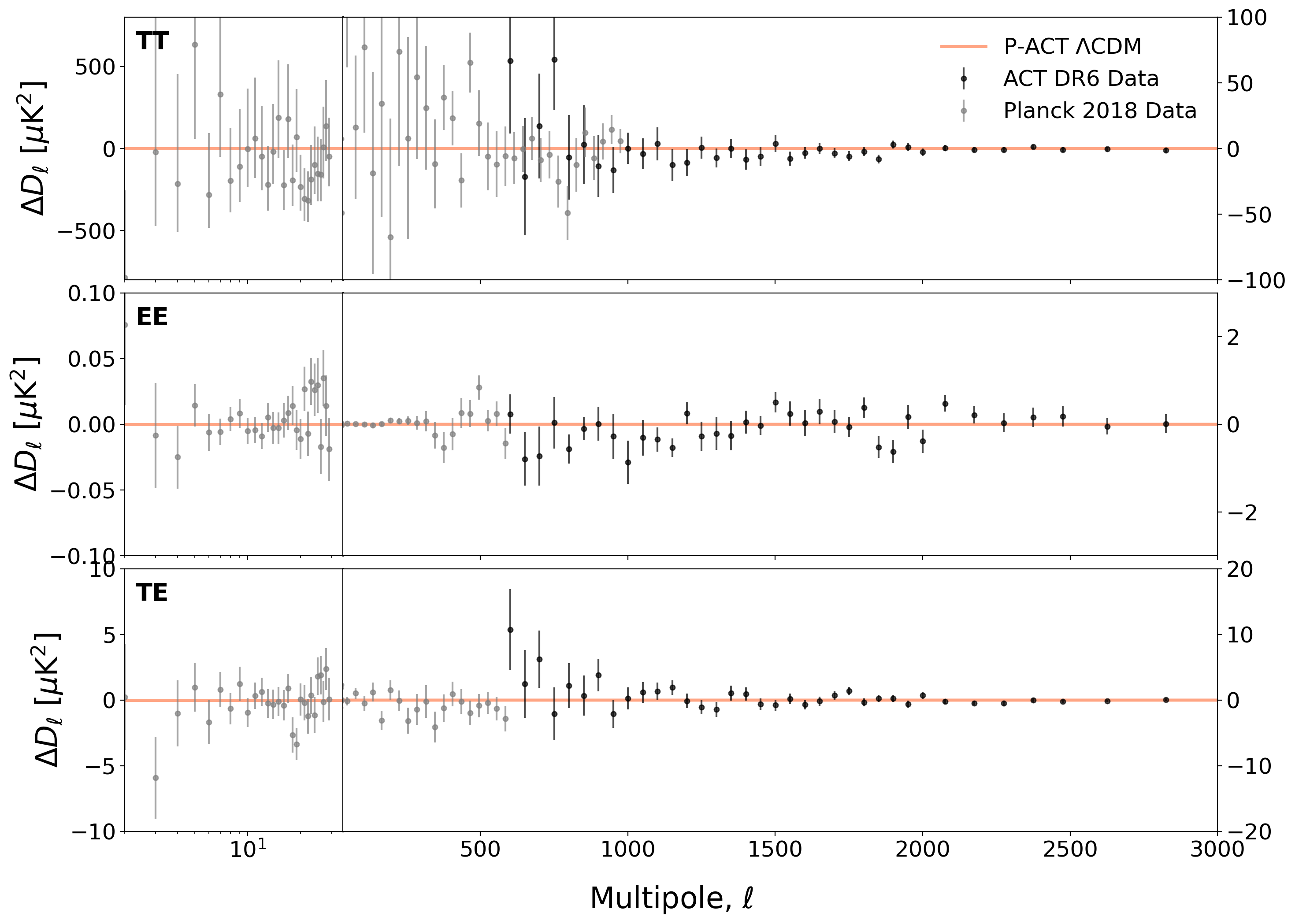}
\caption{Residuals of the CMB power spectra (TT, EE, TE) for \textit{Planck} 2018 (grey) and ACT DR6 (black) data with respect to the P-ACT $\Lambda$CDM best fit (coral). Note that we are only showing the \textit{Planck} data shown is restricted to the P-ACT combination. There are hints of slight oscillatory features around $\Lambda$CDM in these residuals, motivating a study of models that impart oscillatory features.\label{fig: delta cls}} 
\end{figure}

To explore potential departures from the standard power law in a ``model agnostic'' manner, we utilized the model-independent approach we used to obtain constraints for the ACT Data Release 6 (DR6) \citep[][hereafter C25]{ACTextended2025JCAP...11..063C}. While our previous analysis included data from the Dark Energy Spectroscopic Instrument (DESI) Year-1 data \citep{ACTextended2025JCAP...11..063C}, in this work we update the constraints to include the DESI Data Release 2 (DR2) \citep{2025PhRvD.112h3514A, DESIDR22025PhRvD.112h3515A}. Our updated reconstruction can be seen in Figure~\ref{fig:binned Pk} using the priors outlined in Appendix Section~\ref{sec: binned}. There are mild hints of oscillations in the data relative to the $\Lambda$CDM model best fit. In addition, we see these slight oscillations in the residuals of the \textit{Planck} 2018 and ACT DR6 data with $\Lambda$CDM, as can be seen in Figure~\ref{fig: delta cls}. 

Motivated by these hints of oscillations, this paper aims to constrain general oscillatory PPS models as well as oscillations induced by inflationary particle production. While recent work constrained linear and logarithmic oscillating models with ACT, \textit{Planck}, and SPT data \citep{2026PhRvD.113b3544P}. We extend this work by also considering the logarithmic running frequency, and suggest changes to the sampling procedure. We employ a preconditioned sequential Monte Carlo sampler, \texttt{pocoMC} \citep{pocoMCsoft2022JOSS....7.4634K, pocoMCsci2022MNRAS.516.1644K} which is better suited for the multimodal posteriors of these models, rather than simply loosening the Gelman-Rubin convergence criteria \citep{gelman_rubin} to $R - 1 < 0.02$. We discuss the sampler in Section~\ref{ssec: pocomc} and our model selection in Section~\ref{sec:models}. We extend constraints from \textit{Planck} 2018 data on particle production models \citep{Naik2022} with ACT DR6 and DESI DR2 data, as well as extend the $k$-range considered for the oscillatory features. 

Our broad analysis is complemented by recent work by Ref.~\cite{Hidde}, which provides a derivation of the two-point correlation function for a single burst of massive inflationary particle production. They derive and compare constraints from \textit{Planck}, ACT DR6, and DESI DR2 using both Markov Chain Monte-Carlo (MCMC) and matched-filter techniques. While their analysis targets the detailed phenomenology of these isolated bursts, our approach evaluates generalized oscillatory templates and multiple bursts of inflationary particle production. Together, these complementary strategies provide a comprehensive probe of non-standard inflationary dynamics.

This paper is organized as follows. In Section~\ref{sec:models}, we provide an overview of the PPS models we are considering as well as modifications to the particle production model due to numerical instabilities. We discuss the datasets used in this analysis in Section~\ref{sec: data}. In Section~\ref{sec: methodology} we discuss our analysis methodology and integration of \texttt{pocoMC} into \texttt{Cobaya}. We then present our results and discussion in Section~\ref{sec: results} and conclude in Section~\ref{sec: conclusions}.

\section{Models of interest}\label{sec:models}

\subsection{General oscillations} \label{ssec: osc models}

\begin{figure}[htbp!]
    \centering
    \includegraphics[width=0.6\linewidth]{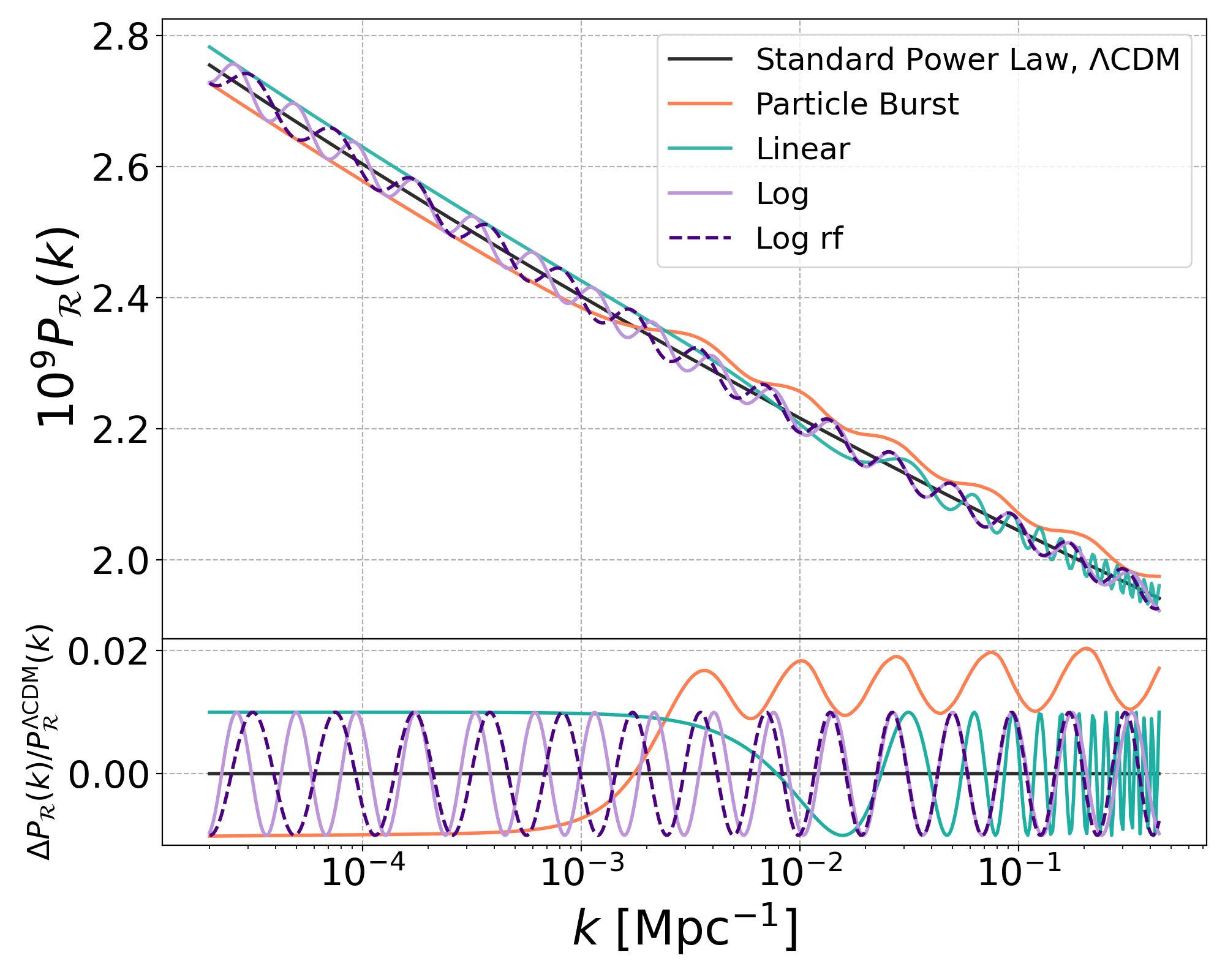}
    \caption{Example primordial power spectra comparing $\Lambda$CDM (dark grey) to Eq.~\ref{eq: oscillating power spec} for $A_X = 0.01$, $\omega_X = 10$, and $\phi_X = 0$ for both linear (teal) and log (light purple) spacing. The logarithmic running model (indigo) uses $\alpha_{rf} = 0.02$. The particle burst model from Eq.~\ref{eq:powerspec multi} (coral) is shown for $A_s = 2.07\times 10^{-9}$, $n_s = 0.965$, $A_I = 5\times 10^{-11}$, $k_1 = 1\times 10^{-3}$, $\Delta= 1.0$, and $i = 10$. The bottom panel shows the fractional residuals with respect to $\Lambda$CDM.}
    \label{fig: power spec ex}
\end{figure}

In order to broadly constrain oscillatory features, we explore general oscillating models as generic templates. Following the \textit{Planck} 2018 inflation paper \citep{2020A&A...641A..10P}, we use,

\begin{align}\label{eq: oscillating power spec}
	P_{\mathcal{R}}(k) &= A_s\bigg(\frac{k}{k_*}\bigg)^{n_s - 1}\Big[1 + \delta P^X\Big], \\
     \delta P^X &= A_X\cos(\omega_X\Xi_X + 2\pi \phi_X),
\end{align}

\noindent where $X = \{\text{lin},\, \text{log},\, \text{log rf}\}$, $\Xi_X = \{k/k_*,\, \ln({k/k_*}),\, \ln({k/k_*})(1 + \alpha_{rf}\ln({k/k_*}))\}$, $A_s$ is the amplitude of the scalar power spectrum, $n_s$ is the scalar spectral index, $k_*$ is the pivot scale, $\omega_X$ is the frequency of oscillations, $\phi_X$ is the phase, and $\alpha_{rf}$ quantifies the running of frequency for the logarithmic model with running frequency. The running frequency model is a good approximation for axion monodromy inflation for $0 \leq \alpha_{\text{rf}} \lesssim 0.01$ \citep{Flauger:2014ana, 2017JCAP...10..058F}. We extend our range for $\alpha_{\text{rf}}$ following Ref. \cite{2020A&A...641A..10P}, allowing $-0.1 \leq \alpha_{\text{rf}} \leq 0.1$. We show example spectra for these models in Figure~\ref{fig: power spec ex}.

\subsection{Particle production during inflation} \label{ssec: part model}

\par
\begin{figure}
\centering
\includegraphics[width=0.4\linewidth]{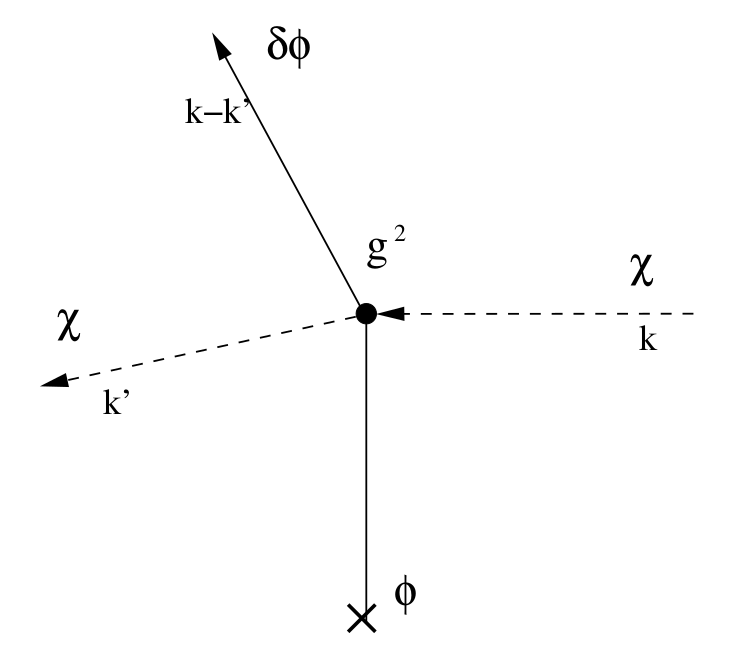}
\caption{Figure 1 from \citep{Barnaby2010}. This diagram shows the rescattering process between the inflaton ($\phi$) and iso-inflaton ($\chi$). This interaction causes Bremsstrahlung radiation of inflation fluctuations, which is the dominant component for the modifications to the PPS. \label{fig:rescattering diagram}} 
\end{figure}

 In the standard inflationary picture, particle production occurs after inflation during preheating and reheating \citep{Dolgov:1982th, 1982PhLB..117...29A}. If there are other subdominant fields coupled to the inflaton, the motion of the inflaton can trigger particle production during inflation \citep{Barnaby2010}. In previous work such as Ref.~\cite{Naik2022}, a generic model is studied, where the inflaton and iso-inflaton interact via
\begin{equation}
    \mathcal{L}_{\text{int}} = -\frac{g^2}{2}\big(\phi - \phi_i\big)^2\chi^2 - g^2\big(\phi - \phi_i\big)\delta\phi\chi^2 - \frac{g^2}{2}\delta\phi^2\chi^2,
\end{equation}
\noindent where $\phi$ is the inflaton, $\phi_i$ corresponds to the point where $\chi$ is instantaneously massless and there will be a burst of particle production, $\chi$ is the iso-inflaton, and $g$ is the dimensionless coupling constant. The primary ways that the $\chi$ particles affect the inflaton are a decrease in velocity after the particles are produced and by rescattering, or in other words Bremsstrahlung radiation of inflaton fluctuations due to rescattering of $\chi$ particles off of the inflaton condensate. The latter effect dominates the signal and therefore was the topic of study of Ref.~\cite{Barnaby2010}. Figure~\ref{fig:rescattering diagram} shows a diagram of this rescattering process. 
\par
\begin{figure}
\centering
\includegraphics[width=0.6\linewidth]{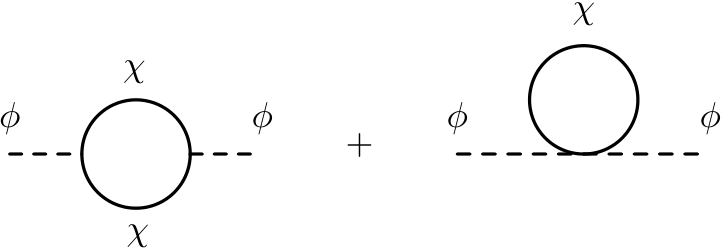}
\caption{Figure 1 from Ref. \citep{1702.07661.2017JCAP...05..054P}, which shows the diagrams depicting the one-loop corrections to the PPS arising from the interactions between the inflaton ($\phi$) and iso-inflaton ($\chi$).} \label{fig: 1 loop rescattering diagram}
\end{figure}

 This effect changes the primordial curvature power spectrum. Ref.~\cite{1702.07661.2017JCAP...05..054P} computed the power spectrum analytically using one-loop approximations, the diagrams for these corrections can be seen in Figure~\ref{fig: 1 loop rescattering diagram}. The PPS is given by:\begin{equation}\label{eq:powerspec multi}
	P_{\mathcal{R}}(k) = A_s\bigg(\frac{k}{k_*}\bigg)^{n_s - 1} + A_I\sum_i\bigg(\frac{f_1(x_i)}{f_1^{\text{max}}}\bigg) + A_{II}\sum_i\bigg(\frac{f2(x_i)}{f_2^{\text{max}}}\bigg),
\end{equation}
\noindent where $A_s$ is the amplitude of the scalar power spectrum, $n_s$ is the scalar spectral index, $k_*$ is the pivot scale, $A_I$ is the amplitude of the particle burst feature and is related to $g$, $k_i$ is the location of this feature and depends on $\phi_i$, and $x_i \equiv \frac{k}{k_i}$. $A_I$ and $A_{II}$ are given by:
\begin{align}
	A_I &\simeq 6.6 \times 10^{-7}g^{7/2}, \\
	A_{II} &\simeq 1.1 \times 10^{-10}g^{5/2} \ln \Big(\frac{g}{0.0003}\Big)^2 \\
    &\simeq (2.9 \times 10^{-6})A_I^{5/7}\Big(\ln A_I^{5/7} + 24\Big), 
\end{align}
\noindent while $f_1(x_i)$ and $f_2(x_i)$ are given by:
\begin{align}
	f_1(x_i) &\equiv \frac{\big(\sin (x_i) - \text{SinIntegral}(x_i)\big)^2}{x_i^3}, \\
	f_2(x_i) &\equiv \frac{-2x_i\cos (2x_i) + (1 - x_i^2)\sin (2x_i)}{x_i^3}, \label{eq: f2}
\end{align}
\noindent and SinIntegral$(x_i) = \int_0^{x_i}\frac{\sin z}{z}dz$ \citep{1702.07661.2017JCAP...05..054P, Naik2022}. We assume ten bursts of particle production as our main particle burst model, which we refer to as ``multiburst'' when comparing to other models. We choose this value of bursts to ensure that the bursts cover our full $k$-range independent of the $k_1$ value within our priors (see Table~\ref{tab:priors}) assuming $\Delta = 1$ (to have minimally overlapping bursts). Figure~\ref{fig: power spec ex} shows an example power spectrum compared to $\Lambda$CDM and the general oscillating models. This model exhibits numerical instabilities specifically from the $f_2(x_i)$ term. In order to alleviate this, we use a Taylor expansion for $f_2(x_i)$ when $x_i < 10^{-5}$, which we discuss in more detail in Appendix Section~\ref{sec: taylor}. 

The separation between the bursts, or bumps in the power spectrum, is parameterized by
\begin{equation}\label{eq: ki}
    k_{i} = e^{(i - 1)\Delta}k_1,
\end{equation}
\noindent where $k_1$ is the initial burst. Note that this assumes that $\Delta$ is constant; Ref.~\citep{Naik2022} similarly assumes that $\Delta$ remains constant under the assumption that for a monomial potential given by\begin{equation} \label{eq: mon pot}
    V(\phi) \propto \frac{\phi^p}{p!},
\end{equation}
\noindent the relative change in $\Delta$ is small, or
\begin{equation}
    \frac{\delta\Delta}{\Delta} \simeq \frac{\delta\phi}{\phi} \lesssim \mathcal{O}(10^{-1}).
\end{equation}
\noindent Large field models such as these are not favoured by current data \citep{2020A&A...641A..10P, ACTextended2025JCAP...11..063C}. Thus, we also consider the so-called E- and T-model $\alpha$-attractor models \citep{2013JCAP...07..002K, 2014arXiv1412.3797G, 2015PhRvD..92f3519C} as they fit well with current data. Additionally, they are useful as generic inflation models, since for certain $\alpha$ values, E- and T-Model potentials are identical to a range of inflationary models including Starobinsky \citep{Starobinsky:1980te} and Higgs inflation \citep{2008PhLB..659..703B}. The E-model potential is given by
\begin{equation}\label{eq:Emodel}
    V\big(\phi\big) = \Lambda^4\bigg(1 - e^{-\sqrt{\frac{2}{3\alpha}}\frac{\phi}{M_{Pl}}}\bigg)^{2},
\end{equation}
while the T-Model potential is specified as
\begin{equation}\label{eq:Tmodel}
    V\big(\phi\big) = \Lambda^4\tanh^{2}{\bigg(\frac{\phi}{\sqrt{6\alpha}M_{Pl}}\bigg)}.
\end{equation}
\noindent Here $\Lambda$ has dimensions of energy, $\phi$ is the inflaton field, and $\alpha$ is a constant that determines the shape of the potential. We use $\alpha = 1$ for this analysis, which corresponds to a tensor-to-scalar ratio value of $r = 3.3\times 10^{-3}$ assuming 60 $e$-folds of inflation, which is well within our current upper bound on $r < 0.032$ at 95\% confidence from \textit{Planck} PR4 and BICEP/Keck \citep{Tristram2022PhRvD.105h3524T}.

Following Ref.~\cite{Naik2022}, we define the $\Delta$ as a function of $\phi$ according to
\begin{equation}
    \Delta(\phi) \simeq \frac{dN}{d\phi}(\phi)2\pi f,
\end{equation}
\noindent where $\phi$ is the inflaton field value, $N$ is the number of $e$-folds, and $f$ is the symmetry breaking scale where particle production takes place when the inflaton crosses the field value of $2\pi fn$, where $n$ takes integer values. The number of $e$-folds as a function of $\phi$ is given by
\begin{equation}
    N(\phi) \simeq \frac{1}{M^2_{Pl}}\int^{\phi}_{\phi_{\text{end}}}d\varphi\frac{V(\varphi)}{V^{\prime}(\varphi)},
\end{equation}
\noindent where $M_{Pl}$ is the reduced Planck mass, $V(\phi)$ is the inflaton potential, $V^{\prime}(\phi)$ is the first $\phi$ derivative of the potential, and $\phi_{\text{end}}$ is where the slow-roll parameter $\epsilon = 1$. The slow-roll parameters are defined as usual:
\begin{equation}
    \epsilon(\phi) = \frac{M^2_{Pl}}{2}\Bigg(\frac{V^{\prime}(\phi)}{V(\phi)}\Bigg)^2 \qquad, \qquad \eta(\phi) = M^2_{Pl}\frac{V^{\prime\prime}(\phi)}{V(\phi)}.
\end{equation}

\begin{figure}[htbp!]
\centering
\includegraphics[width=0.9\linewidth]{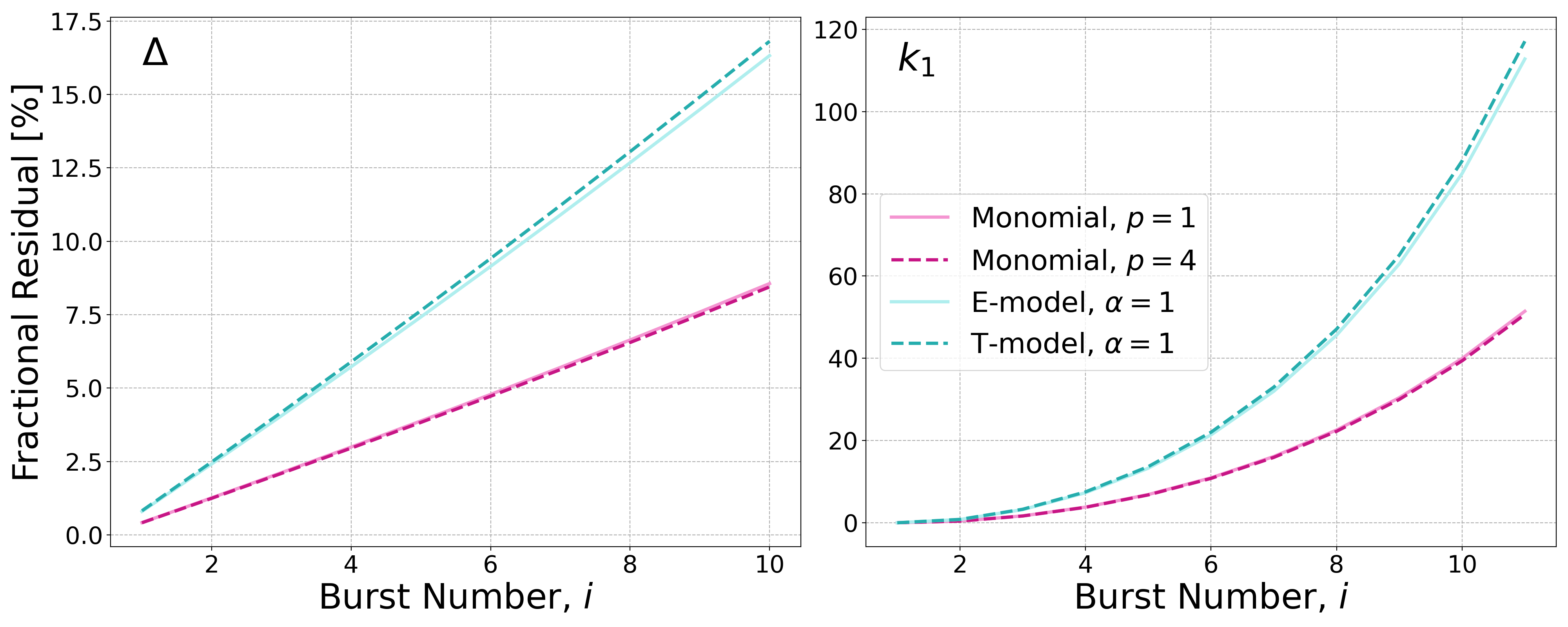}
\caption{Relative shifts in the burst spacing parameter $\Delta$ (left) and the resulting burst locations $k_i$ (right) for 10 consecutive bursts, comparing the exact calculation to an assumption of constant $\Delta$. We show the results for the monomial potentials for $p = 1$ (light pink) and $p = 4$ (dark pink) as well as the E-model (light turquoise) and T-model (dark turquoise) for $\alpha = 1$. Note that the drift in $k_i$ is cumulative and thus results in a significant difference of up to 51.4\% and 117\% for the monomial and $\alpha$-attractor potentials, respectively.} \label{fig: delta shift}
\end{figure}

Similarly to \cite{Naik2022}, we define the change in $\Delta$ in terms of the change in field values and slow-roll parameters, or 
\begin{equation}
    \frac{\delta\Delta}{\Delta} \simeq \sqrt{2\epsilon}\Big(1 - \frac{\eta}{2\epsilon}\Big)\frac{\delta\phi}{M_{Pl}}.
\end{equation}
\noindent We then determine what the relative change in $\Delta$ is for 1 -- 10 bursts of particle production. We consider the monomial potential from Eq.~\ref{eq: mon pot}, for $p = 1, 4$ as example exponents, as the results do not have a strong dependence on the exponent choice, as well as the E- and T-model $\alpha$-attactors. The results for both the drift in $\Delta$ and $k_i$ can be seen in Figure~\ref{fig: delta shift}. The drift effect is low for a small number of bursts, but the difference in $\Delta$ for 10 bursts is 8.6 (8.4)\%  and 16.3 (16.8)\% for the monomial potential for $p =$ 1 (4) and $\alpha$-attractor E-model (T-model), respectively. This effect is compounded as it is cumulative for $k_i$, and we see a total shift for 10 bursts of 51.4 (50.7)\% and 113 (117)\% for the monomial potential for $p =$ 1 (4) and $\alpha$-attractor E-model (T-model), respectively. Given that the $k_1$ values preferred by our data are large ($k_1 \simeq 10^{-1}$) for our multiburst model over the entire $k$-range, we do not have many bursts within our $k$-range and thus still assume a constant $\Delta$. Allowing for variation in $\Delta$ could generate large effects for realistic inflaton models and a large number of bursts; we leave this investigation to future work.
\section{Datasets considered} \label{sec: data}
We use the ACT DR6 \texttt{ACT-lite} CMB-only foreground-marginalized likelihood as described in Refs.~\cite{ACTlcdm2025JCAP...11..062L, ACTextended2025JCAP...11..063C}. These temperature-temperature power spectra (TT), temperature-E mode polarization power spectra (TE), and polarization-polarization power spectra (EE) are derived from maps made from observations from 2017 -- 2022 using three frequency bands: ``f090" (77--112 GHz), ``f150'' (124 -- 172 GHz), and ``f220'' (182 -- 277 GHz). For large-scale polarization constraints we use the \textit{Planck} low-multipole ``Sroll2'' EE power spectrum likelihood \citep{Sroll22020A&A...635A..99P}. Our \textit{Planck} data constraints are from the low-$\ell$ TT likelihood as well as the TT, TE, and EE \texttt{plik\_lite} PR3 likelihood \citep{planckoverview2020A&A...641A...1P}. When we are combining \textit{Planck} and ACT (P-ACT), we use the ``Planck$_{\mathrm{cut}}$" likelihood which truncates the multipole range to $\ell < 1000$ for TT and $\ell < 600$ for TE/EE as described in C25. When considering ACT data on its own, we apply the Sroll2 Gaussian prior $\tau_\mathrm{reio} = 0.0566 \pm 0.0058$. 

In addition to the CMB primary anisotropy data, we use the combined ACT DR6 CMB lensing likelihood \citep{QU2024ApJ...962..112Q, madhavacheril2024ApJ...962..113M} and \textit{Planck} PR4 NPIPE lensing likelihood\citep{NPIPE2022JCAP...09..039C} as described in C25. Finally, as a probe of the overall expansion rate from large-scale structure, we use the DESI Baryon Acoustic Oscillation (BAO) DR2 likelihood \citep{DESIDR22025PhRvD.112h3515A}. The combination of \textit{Planck}, ACT DR6, lensing, and DESI DR2 is called ``P-ACT-LB'' throughout this paper, other than for our comparison with the binned PPS reconstruction where we call it ``P-ACT-LB$_{\text{DR2}}$" to avoid confusion with the original results with DESI Year 1.
\section{Analysis methodology} \label{sec: methodology}

\subsection{Overview and priors}

We use modified versions of the Code for Anisotropies in the Microwave Background \citep[\texttt{CAMB},][]{CAMB2011ascl.soft02026L} for our primordial power spectra models outlined in Equations~\ref{eq: oscillating power spec} and \ref{eq:powerspec multi}. We use the \texttt{CAMB} accuracy settings described in Appendix A of C25. For the general oscillation models, we also implement the non-linear matter power spectrum form for oscillatory primordial features outlined in Ref.~\cite{WMP2019PhRvR...1c3209B}; see Appendix Section~\ref{ssec: matter power} for more information. The priors we use are specified in Table~\ref{tab:priors}. We use the Preconditioned Monte Carlo algorithm \citep[\texttt{pocoMC},][]{pocoMCsoft2022JOSS....7.4634K, pocoMCsci2022MNRAS.516.1644K} to get our full posteriors and the \texttt{Cobaya} \citep{cobaya2021JCAP...05..057T} minimizer (setting \texttt{best\_of:10}) to find the maximum \textit{a posteriori} (MAP) points for each model. For the updated binned P-ACT-LB$_{\text{DR2}}$ we use the traditional MCMC sampler within \texttt{Cobaya} and the priors defined in C25 and Appendix Section~\ref{sec: binned} in order to be consistent with our previous analyses.

\begin{table}
\centering
\begin{tabular}{l|c}
\toprule
\textbf{Parameter} & \textbf{Prior} \\ 
\midrule
\multicolumn{2}{l}{\textbf{$\mathbf{\Lambda}$CDM}} \\ 
\midrule
$\Omega_\mathrm{b}h^2$ & $[0.02, 0.024]$ \\ 
$\Omega_\mathrm{c}h^2$ & $[0.1, 0.14]$ \\ 
$\theta_\mathrm{MC}$ & $[0.0103, 0.0105]$ \\ 
$\log(10^{10} A_\mathrm{s})$ & $[2.9, 3.2]$ \\ 
$n_\mathrm{s}$ & $[0.92, 1.02]$ \\ 
$\tau_\mathrm{reio}$ & $[0.01, 0.1]$ \\ 
\midrule
\multicolumn{2}{l}{\textbf{ACT Specific}} \\ 
\midrule
$A_{\rm ACT}$ & $\mathcal{N}(1.0,0.003)$ \\
$p_{\rm ACT}$ & $[0.5, 1.5]$ \\ 
\midrule
\multicolumn{2}{l}{\textbf{General Oscillations}} \\ 
\midrule
$A_{X}$ & $[0, 0.5]$ \\ 
$\log_{10}(\omega_{X})$ & $[0, 2]$ \\
$\phi_{X}$ & $[0, 1]$ \\ 
$\alpha_{\mathrm{rf}}$ & $[-0.1, 0.1]$ \\ 
\midrule
\multicolumn{2}{l}{\textbf{Particle Bursts}} \\ 
\midrule
$\log(10^{12} A_I)$ & $[-2.3, 8]$ \\ 
$\log_{10}(k_1)$ & $[-4.5, -0.365]$ \\ 
$\Delta$ & $[0, 3]$ \\ 
\bottomrule
\end{tabular}
\caption{Priors for all parameters. The priors are flat unless otherwise specified. For the ACT-alone runs, we instead use the Sroll2 Gaussian prior: $\tau_\mathrm{reio}$:  $\mathcal{N}(0.0566,0.0058)$}
\label{tab:priors}
\end{table}
\subsection{\texttt{pocoMC} and integration into \texttt{Cobaya}} \label{ssec: pocomc}

Sampling these alternative PPS models requires efficient sampling of high-dimensional, often highly degenerate, and multimodal posterior distributions. This is a known issue for traditional MCMC methods, leading to prohibitive run times that can hinder model exploration. To alleviate this problem, we integrated the \texttt{pocoMC} machine-learning-based Preconditioned Monte Carlo sampler into \texttt{Cobaya}. \texttt{pocoMC} utilizes a normalizing flow to decorrelate parameters for each temperature step in an adaptive Sequential Monte Carlo sampling scheme of the form \citep{pocoMCsci2022MNRAS.516.1644K}:
\begin{equation}
p_t(\theta) \propto \pi(\theta)\mathcal{L}(\theta)^{\beta_t}, \quad t = 1, \dots, T, 
\end{equation}
\noindent where $\pi(\theta)$ is the prior, $\mathcal{L}(\theta)$ is the likelihood, and $\beta$ is the inverse temperature parameter given by
\begin{equation}
\beta_1 = 0 < \beta_2 < \cdots < \beta_T = 1.
\end{equation}
\noindent The change in annealing temperature is computed adaptively in order to have a constant number of effective particles at each step. The normalizing flow is used to learn the mapping between the target distribution $p_t(\theta)$ and a Gaussian. This development is crucial as it is model independent and learns the target posterior in real time. In contrast, many other machine-learning-based sampling methods require a significant amount of model-specific pre-training before sampling. At each temperature step, a $t$-preconditioned Crank-Nicolson MCMC algorithm \citep{tpcn2024InvPr..40l5023G} is used to sample the space. See Appendix Section~\ref{ssec: settings} for the \texttt{pocoMC} settings we used. By integrating \texttt{pocoMC} into \texttt{Cobaya}, we can easily apply the sampler to existing CMB theory codes and datasets, which are already formatted for use within \texttt{Cobaya}. 

\section{Results and discussion} \label{sec: results}
\subsection{General oscillations}
There are a few variations of the general oscillation models that are parameterized with amplitudes and spectral indices in Eq.~\ref{eq: oscillating power spec} and differing assumptions on the spacing of the model in wavenumber, which changes the support of the oscillations in the primordial power across scales.
\begin{figure}[htbp!]
    \centering
    \includegraphics[width=0.9\linewidth]{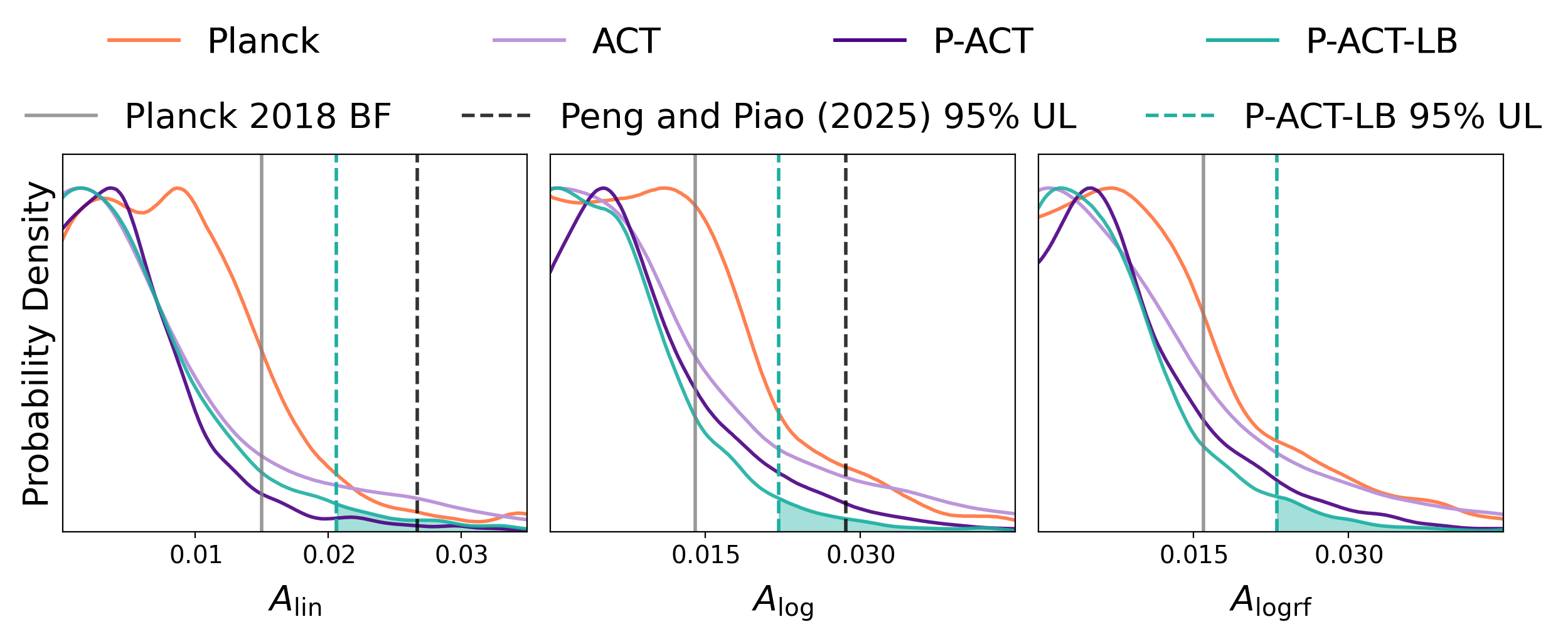}
    \caption{1D posterior distributions for the amplitude for the linearly spaced (left), logarithmically spaced (middle), and logarithmically spaced with a running frequency (right) oscillating $P_{\mathcal{R}}(k)$ models outlined in Eq.~\ref{eq: oscillating power spec}. We show the results for \textit{Planck} (coral), ACT (light purple), P-ACT (indigo), and P-ACT-LB (teal). The best-fit (BF) results from \textit{Planck} \citep{2020A&A...641A..10P} are shown (grey) and the 95\% upper limits (UL) from \textit{Planck}+SPT+ACT \citep{2026PhRvD.113b3544P} (dark grey) are shown as dashed lines.}
    \label{fig: Aocs 1d}
\end{figure}
\subsubsection{Amplitude of oscillations and spectral index}
The allowed fraction of oscillating power in addition to the usual power law is constrained to be less than $\sim2\%$ at 95\% confidence for all models considered. We show the 1D posterior distributions for the amplitude in Figure~\ref{fig: Aocs 1d} with the best-fit (BF) values from the \textit{Planck} 2018 analysis \citep{2020A&A...641A..10P} and the 95\% upper limits (UL) from the MCMC analysis using \textit{Planck}+SPT+ACT \citep{2026PhRvD.113b3544P}. Using the P-ACT-LB constraints, we obtain 95\% upper limits on the amplitudes of
\begin{eqnarray}
A_{\text{lin}} & < & 0.021 \ \ (95\% \text{ CL}), \nonumber \\
A_{\text{log}} & < & 0.022 \ \ (95\% \text{ CL}), \nonumber \\
A_{\text{log rf}} & < & 0.023 \ \ (95\% \text{ CL}). \nonumber 
\end{eqnarray}
\noindent See Table~\ref{tab: A UL} for the upper limits for all datasets. This bound, which restricts the additional oscillatory power to less than around 2\% of the scalar non-oscillatory signal, represents a marginal improvement over the constraints from Ref.~\cite{2026PhRvD.113b3544P} of $\sim2.9\%\, A_s$, this is likely due to the choice of sampling methods (see Section~\ref{ssec: pocomc} for some description of samplers in this work). Similarly, we see improvements ranging from 40 -- 57\% in the allowed amplitude when comparing our MAP amplitude values to the \textit{Planck} 2018 best-fit values from Ref.~\cite{2020A&A...641A..10P}, which is likely driven by the inclusion of the ACT DR6 high-$\ell$ data. In line with these previous analyses, we do not see a large variation in our amplitude constraints between the three different oscillation spacing models considered. 

\begin{table}
\centering
\begin{tabular}{l|c|c|c|c}
\toprule
\textbf{Parameter} & \textbf{\textit{Planck}} & \textbf{ACT} & \textbf{P-ACT} & \textbf{P-ACT-LB} \\ 
\midrule
$A_{\text{lin}}$ & $< 0.023$ & $< 0.033$ & $< 0.017$ & $< 0.021$ \\ 
$A_{\text{log}}$ & $< 0.032$ & $< 0.040$ & $< 0.027$ & $< 0.022$ \\ 
$A_{\text{log rf}}$ & $< 0.034$ & $< 0.040$ & $< 0.027$ & $< 0.023$ \\ 
\bottomrule
\end{tabular}
\caption{The 95\% CL upper limits on the general oscillating model amplitudes. We show the results for \textit{Planck}, ACT, P-ACT, and P-ACT-LB.}
\label{tab: A UL}
\end{table}
\begin{figure}
    \centering
    \includegraphics[width=0.9\linewidth]{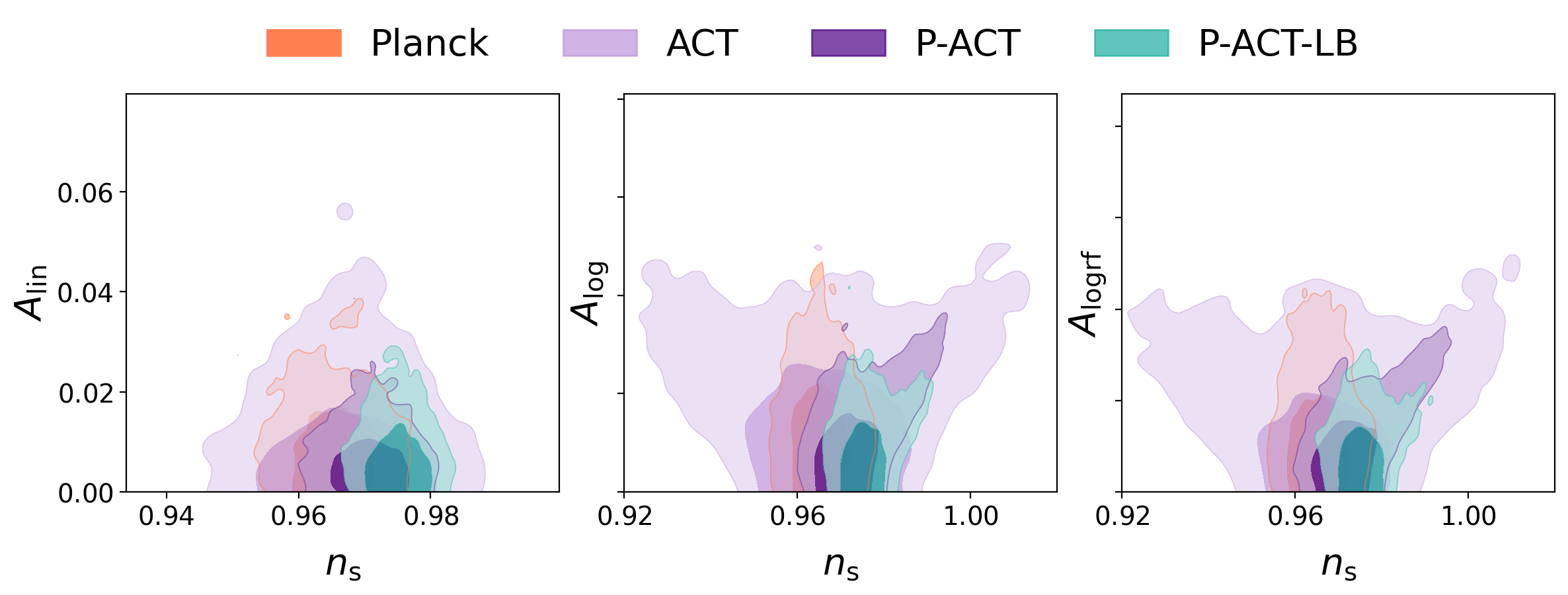}
    \caption{2D posterior distributions of $n_s$ and $A_\mathrm{{osc}}$ for the general oscillating models for \textit{Planck} (coral), ACT (light purple), P-ACT (indigo), and P-ACT-LB (teal). The shift in $n_s$ is also present in the $\Lambda$CDM constraints and is not a new feature of this model. We do not see a significant shift between the logarithmically spaced and logarithmically spaced with a running frequency, likely due to the running being unconstrained.}
    \label{fig: ns Aosc}
\end{figure}
The tightening in the constraints of the amplitude of oscillations has different implications depending on the model being considered. For example, if we were to relate the constraints on $A_{\text{log rf}}$ to axion monodromy inflation (the logarithmic running frequency model is a good approximation for the PPS of this model) we can infer constraints on the field value at the time the pivot scale exits the horizon and the axion decay constant, see Refs.~\citep{2010JCAP...06..009F, Flauger:2014ana}. Figure 1 in Ref. \cite{2010JCAP...06..009F} shows the relationship between the amplitude and decay constant computed both analytically and numerically. There is a general trend that for a small decay constant, you correspondingly have a small amplitude (until the decay constant $\gtrsim 0.02$ where the amplitude does decrease marginally for a larger decay constant).

\begin{figure}[htbp!]
    \centering
    \includegraphics[width=0.6\linewidth]{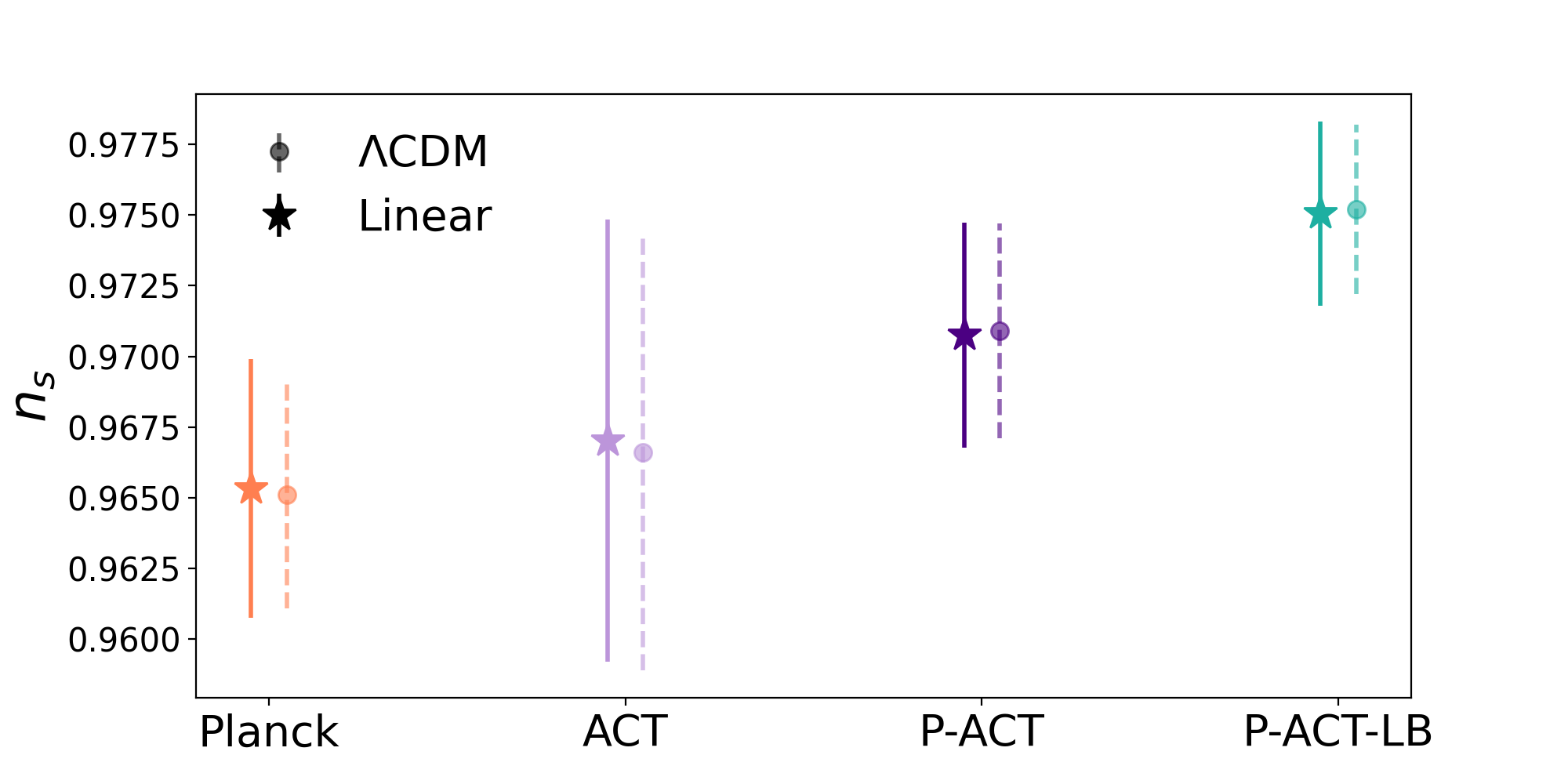}
    \caption{1D marginalized 68\% CL on $n_s$ for the linear oscillating model (stars) and $\Lambda$CDM values from Ref. \cite{ACTlcdm2025JCAP...11..062L} (circles) for \textit{Planck} (coral), ACT (light purple), P-ACT (indigo), and P-ACT-LB (teal). Note that the shift to higher $n_s$ that we see is consistent with the trend observed in ACT DR6 constraints when only considering the $\Lambda$CDM model.}
    \label{fig: ns}
\end{figure}

We present the 2D posterior distribution for $n_s$ and $A_X$ for both the linear and logarithmic models in Figure~\ref{fig: ns Aosc}. Similar to the 1D limits, there is no significant deviation from the logarithmic results when a running frequency is included. This is likely due to $\alpha_{\text{rf}}$ being unconstrained (and hence the MAP being close to zero); there is no strong preference for running from the data, as shown in Table~\ref{tab: osc_comparison} and Figure~\ref{fig: logrf all} in Appendix Section~\ref{ssec: full osc}. There is, however, a shift in $n_s$ when moving from \textit{Planck} to ACT and P-ACT, with the largest shift for P-ACT-LB. This shift was seen in the $\Lambda$CDM best-fit values presented in Ref. \cite{ACTlcdm2025JCAP...11..062L}, as also shown in Figure~\ref{fig: ns}, where we plot the 1D marginalized 68\% CL on $n_s$ for the linear oscillating model and $\Lambda$CDM.
Our values for $n_s$ in these general oscillating models are consistent with the $n_s$ values obtained when assuming the $\Lambda$CDM cosmology. The result of our model flexibility is only that the errors on $n_s$ increase slightly; the main impact of our models is changing the \textit{amplitude} of oscillations and not the spectral index. While we focused on the most relevant parameter constraints in this section, the full set of parameter constraints are provided for completeness in Appendix Section~\ref{ssec: full osc}. 

\subsubsection{Phase shift between \textit{Planck} and P-ACT}
\begin{figure}
    \centering
    \includegraphics[width=0.9\linewidth]{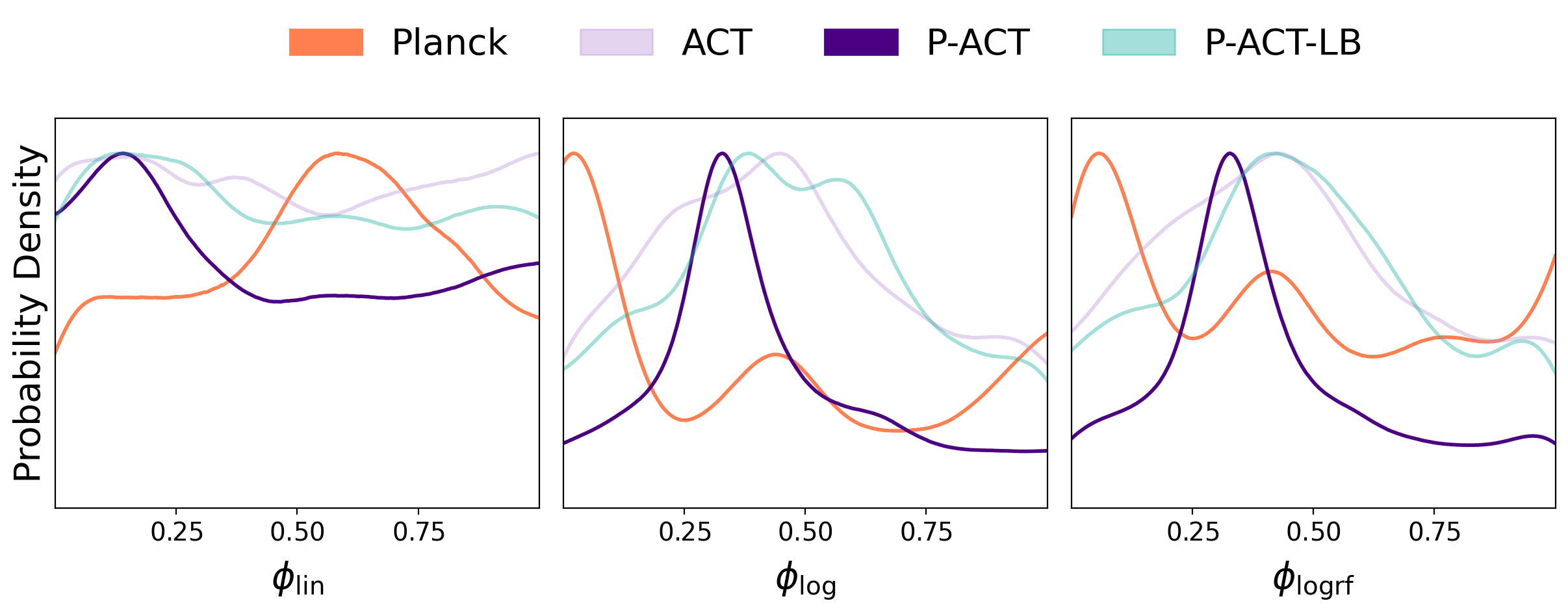}
    \caption{1D posterior distributions for the phase of the linearly spaced (left), logarithmically spaced (middle), and logarithmically spaced with a running frequency (right) oscillating models. We show the results for \textit{Planck} (coral), ACT (light purple), P-ACT (indigo), and P-ACT-LB (teal). We see a difference in the preferred phase between \textit{Planck} and P-ACT which is consistent with our binned PPS reconstruction findings.}
    \label{fig: phi 1d}
\end{figure}
\begin{figure}
    \centering
    \includegraphics[width=0.9\linewidth]{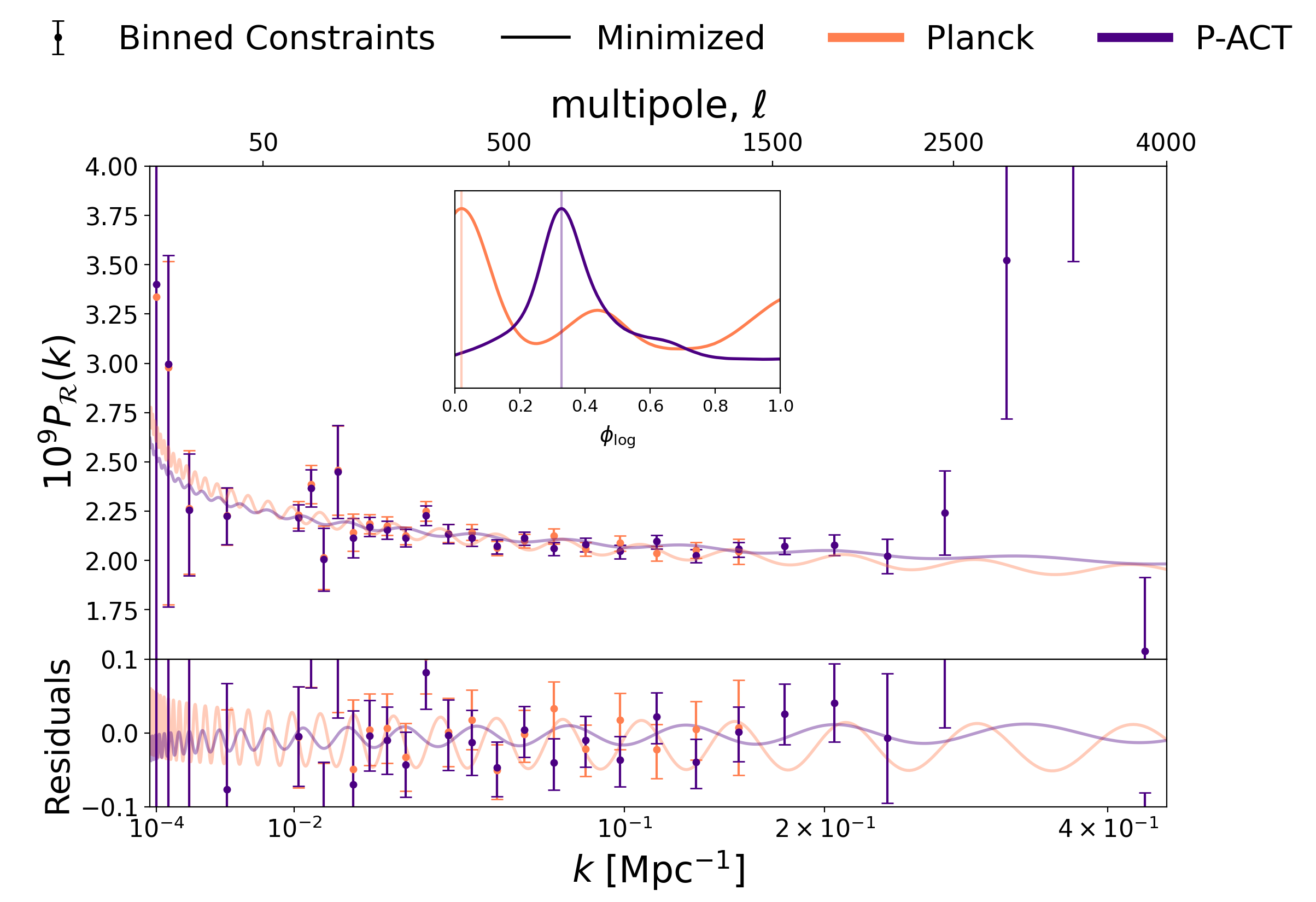}
    \caption{Comparing the phase shift for \textit{Planck} and P-ACT. Our binned $P_{\mathcal{R}}(k)$ constraints from Ref.~\cite{ACTextended2025JCAP...11..063C} are shown as error bars. The 1D posterior for the phase of the logarithmically spaced model is shown with a vertical line indicating the peak of the distributions. The results of running a minimizer with the phase set to these values are shown as faint lines. The residuals are with respect to the best-fit $\Lambda$CDM PPS for the respective dataset. These MAP values have resulting $\Delta\chi^2$ of $-10.14$ and $-2.27$ for \textit{Planck} and P-ACT, respectively.}
    \label{fig: phase diff}
\end{figure}

We previously observed a slight phase shift in the data for \textit{Planck} compared to P-ACT in C25, which can be seen in the residuals of Figure~\ref{fig:binned Pk}, where the binned PPS reconstructed from \textit{Planck} data are mildly out of phase compared to the P-ACT data. We observe a similar trend when considering the constraints on the phase ($\phi_X$) in Eq.~\ref{eq: oscillating power spec}, of the general oscillating models for \textit{Planck} and P-ACT. This can be seen in Figure~\ref{fig: phi 1d} which shows the 1D posterior distributions for $\phi_X$. In order to further test this result, we fixed $\phi_{\text{log}}$ to the peak value for both \textit{Planck} ($\phi_{\text{log}} = 0.0215$) and P-ACT ($\phi_{\text{log}} = 0.328$). We then used the \texttt{Cobaya} minimizer to find the MAP spectra for these fixed phase shift values. Figure~\ref{fig: phase diff} shows both the 1D posterior distributions for $\phi_{\text{log}}$ with the peak values indicated with vertical lines and the resulting MAP spectra and corresponding binned results. Our corresponding $\Delta\chi^2 = \chi^2_{\text{model}} - \chi^2_{\Lambda\text{CDM}}$ values are $-10.14$ and $-2.27$ for \textit{Planck} and P-ACT, respectively. The MAP spectra exhibit agreement with the binned results, with the \textit{Planck} MAP spectra also having a higher frequency of oscillations compared to the P-ACT map spectra, to more closely follow the \textit{Planck} data. While we do not present Bayesian evidence values for these specific phase shift investigations and MAP spectra, as will be seen in Section~\ref{ssec: model comparison}, all oscillating models for which we compute the Bayesian evidence show a moderate preference for $\Lambda$CDM.

\subsubsection{Frequency of oscillations}

\begin{figure}
    \centering
    \includegraphics[width=0.9\linewidth]{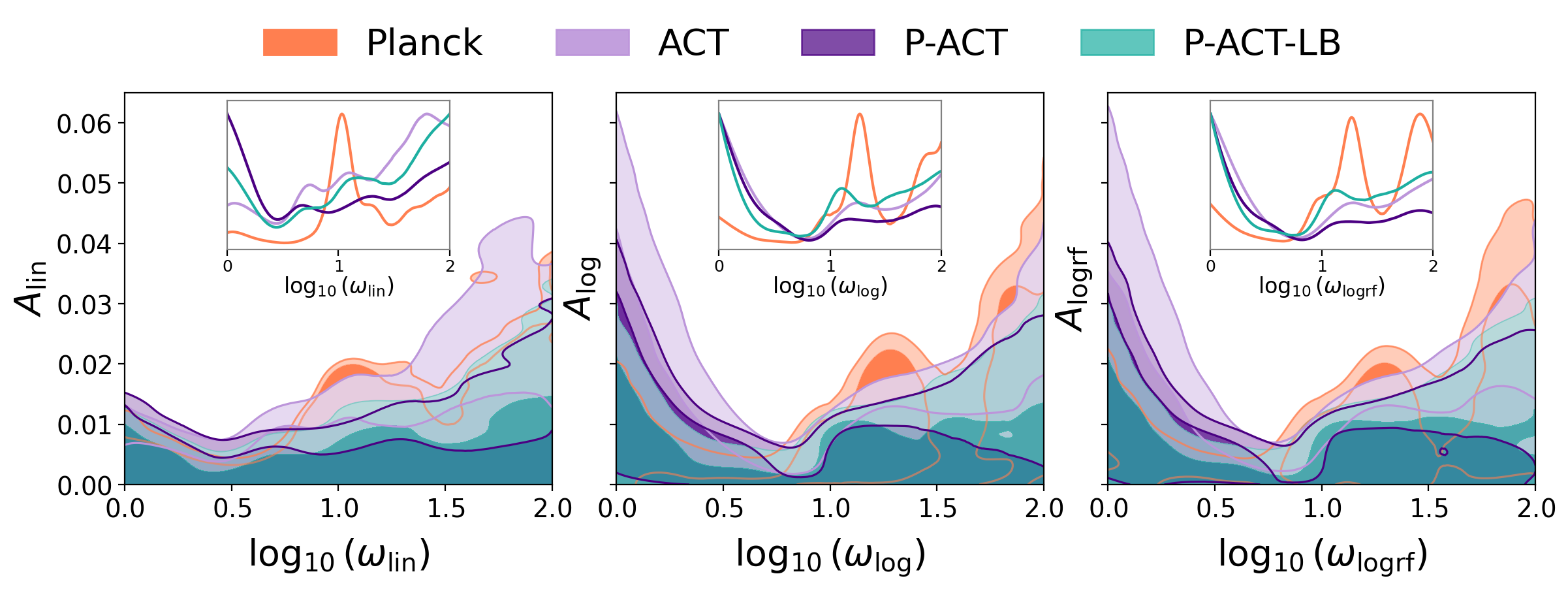}
    \caption{2D posterior distributions for the amplitude and frequency for the linearly spaced (left), logarithmically spaced (middle), and logarithmically spaced with a running frequency (right) oscillating models. The 1D posterior distributions for the frequency are also shown for each model. We show the results for \textit{Planck} (coral), ACT (light purple), P-ACT (indigo), and P-ACT-LB (teal).}
    \label{fig: Aocs and logogema}
\end{figure}

The implications of constraining the frequency depend on the specific model in question. But, for example, if we consider a particle production model mapped onto a general oscillation model such as Eq.~\ref{eq: oscillating power spec}  (as opposed to the more specific formulation we study in Eq.~\ref{eq:powerspec multi} and present in Section~\ref{ssec: part constraints}), the spacing $\Delta$ between bumps in the PPS can be related to a frequency of oscillations, which corresponds to the number of $e$-folds of expansion between bursts of particle production. We present our frequency constraints in Figure~\ref{fig: Aocs and logogema}, which shows the 2D posterior distributions for $A_X$ and $\log_{10}(\omega_X).$ The inset 1D distributions in the Figure show the 1D posterior distributions for the frequency alone. Our results for $\log_{10}(\omega_X)$ are fairly unconstrained and multimodal which is consistent with previous analyses \cite{2020A&A...641A..10P, 2026PhRvD.113b3544P}. We do, however, see a general trend for a larger frequency for \textit{Planck} compared to the other datasets. Additionally, for the data combinations including ACT, small values of $\log_{10}(\omega_X)$ are preferred for large amplitudes, whereas the \textit{Planck} data show no preference for a small frequency (those $\log_\mathrm{10}(\omega_X) \lesssim 1$) regardless of amplitude. Despite this trend in the 2D distributions, when we compute the MAP spectra considering all parameters, we get very similar values for $\log_{10}(\omega_X)$ except for logarithmic and running models for ACT alone, which prefers larger frequency values, as can be seen in Table~\ref{tab: osc_comparison}. 

We can do a crude comparison of the MAP frequency we obtain for the logarithmically spaced oscillations and the particle production model with a constant $\Delta$. We consider the spacing between oscillations:
\begin{equation}
    \omega_{\text{log}}\ln\bigg(\frac{k_{i+1}}{k_*}\bigg) - \omega_{\text{log}}\ln\bigg(\frac{k_{i}}{k_*}\bigg) = 2\pi.
\end{equation}
\noindent We can substitute this relationship for the spacing in the particle burst model, $k_{i+1} = e^{\Delta}k_i$, which gives,
\begin{equation}
    \omega_{\text{log}}\ln\big(e^{\Delta}\big) = 2\pi \quad \implies \quad \Delta = \frac{2\pi}{\omega_{\text{log}}}.
\end{equation}

Using the MAP constraint for P-ACT-LB on the frequency of oscillations, we obtain an equivalent $\Delta \simeq 0.55$. Similar to the frequency constraints here and the effective mapping, $\Delta$ is fairly unconstrained in our specific particle burst model analysis, as discussed in Section~\ref{ssec: part constraints}, however, we get a MAP value of $\Delta = 0.785$ for P-ACT-LB. This value differs from the more simple mapping above as might be expected: unlike the general oscillation models described here, the specific particle burst model also varies the initial scale, $k_1$, at which the bumps appear. The values obtained from these general templates can however be useful to get a broad overview of what models may fit the data, but as always, each model of interest should be individually studied to derive specific constraints.

\begin{table*}[htbp!]
\centering
\begin{adjustbox}{max width=\textwidth} 
\begin{tabular}{l|cccc|cccc|cccc}
\toprule
 & \multicolumn{4}{c}{\textbf{Linear}} & \multicolumn{4}{c}{\textbf{Log}} & \multicolumn{4}{c}{\textbf{Log rf}} \\
\cline{2-13}
\textbf{Parameter} & \textbf{\textit{Planck}} & \textbf{ACT} & \textbf{P-ACT} & \textbf{P-ACT-LB} & \textbf{\textit{Planck}} & \textbf{ACT} & \textbf{P-ACT} & \textbf{P-ACT-LB} & \textbf{\textit{Planck}} & \textbf{ACT} & \textbf{P-ACT} & \textbf{P-ACT-LB} \\
\midrule
$\log(10^{10} A_\mathrm{s})$ & 3.06 & 3.06 & 3.05 & 3.06 & 3.05 & 3.05 & 3.06 & 3.07 & 3.05 & 3.06 & 3.05 & 3.06 \\ 
$n_\mathrm{s}$ & 0.965 & 0.965 & 0.970 & 0.975 & 0.964 & 0.968 & 0.971 & 0.977 & 0.964 & 0.968 & 0.971 & 0.976 \\
$10^3A_{X}$ & 11.7 & 9.57 & 7.91 & 7.93 & 8.99 & 13.7 & 5.55 & 8.18 & 10.7 & 11.9 & 9.03 & 6.96 \\
$\log_{10}(\omega_{X})$  & 1.06 & 0.972 & 1.08 & 1.08 & 0.994 & 1.25 & 1.05 & 1.06 & 1.01 & 1.29 & 1.05 & 1.05 \\
$\phi_{X}$ & 0.447 & 0.394 & 0.439 & 0.510 & 0.436 & 0.505 & 0.681 & 0.643 & 0.486 & 0.459 & 0.577 & 0.540 \\
$\alpha_{\mathrm{rf}}$ & -- & -- & -- & -- & -- & -- & -- & -- & 0.0278 & -0.0733 & 0.0274 & 0.0506 \\
\midrule
$\Delta\chi^2$ & $-8.93$ & $-3.55$ & $-4.68$ & -5.53 & $-6.49$ & $-5.26$ & $-6.15$ & -7.18 & $-4.98$ & $-6.83$ & $-7.69$ & -8.70 \\
$\ln B$ & $-2.60$ & $-2.94$ & $-2.97$ & $-3.84$ & $-2.71$ & $-2.91$ & $-2.81$ & $-3.49$ & $-2.73$ & $-2.95$ & $-2.88$ & $-3.58$ \\
\bottomrule
\end{tabular}
\end{adjustbox}
\caption{MAP values for the general oscillation models as well as $\Delta\chi^2$ and $\ln B$ relative to $\Lambda$CDM. Negative $\Delta\chi^2$ indicates an improved fit and negative $\ln B$ indicates a preference for $\Lambda$CDM. The $\Delta\chi^2$ values are computed using the MAP values, whereas the $\ln B$ values are computed using the evidence value from \texttt{pocoMC}.}
\label{tab: osc_comparison}
\end{table*}

\subsection{Particle production during inflation} \label{ssec: part constraints}
\begin{figure}[htbp!]
    \centering
    \includegraphics[width=0.6\linewidth]{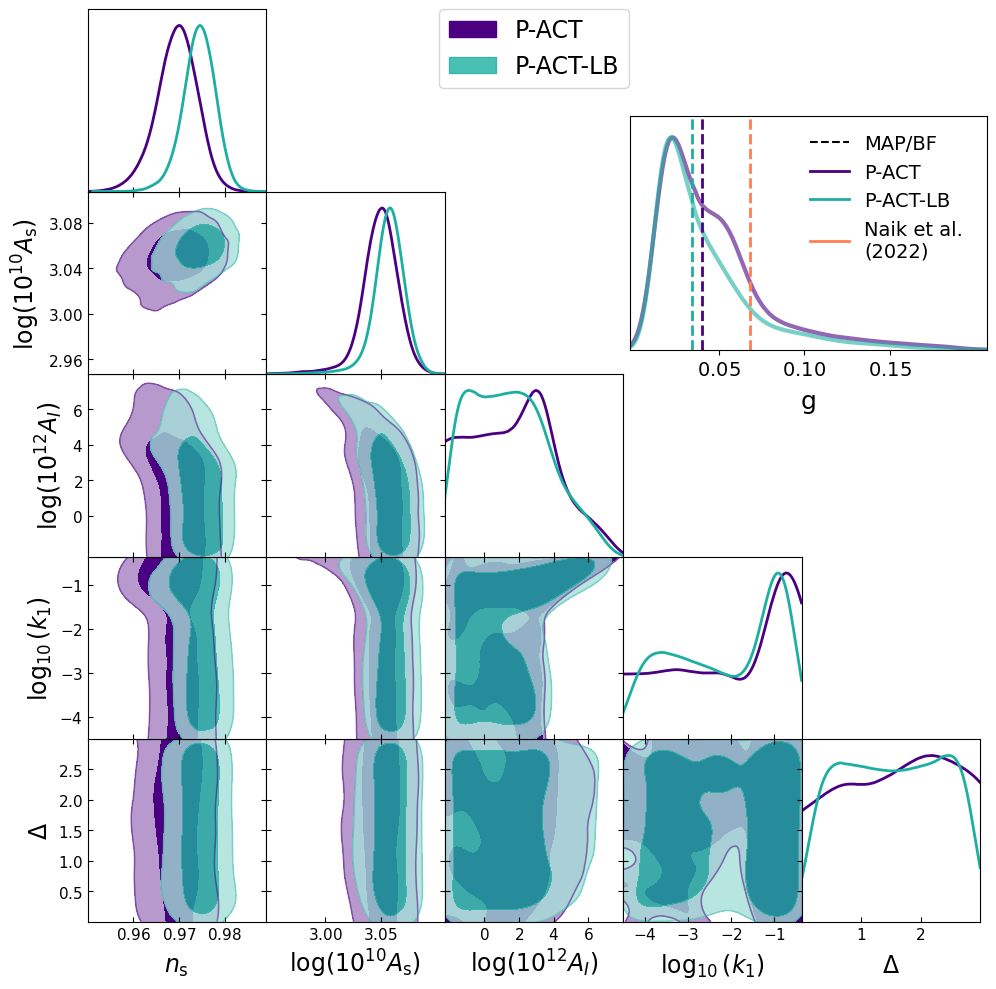}
    \caption{1D and 2D marginalized posterior distributions for the multiburst particle production model outlined in Eq.~\ref{eq:powerspec multi} for both P-ACT (indigo) and P-ACT-LB (teal). The 1D posterior distributions for the amplitude and derived coupling constant is shown with the MAP values as dashed lines for our results as well as the best-fit (BF) value from Ref. \cite{Naik2022} (coral dashed line).}
    \label{fig:triangle multi A k Delta}
\end{figure}
We now discuss constraints on inflationary particle production, focusing on specific parameters of interest. The full results and parameter correlations are presented in Appendix Section~\ref{ssec: full part} for completeness. In Figure~\ref{fig:triangle multi A k Delta} we plot the posterior distributions for both P-ACT and P-ACT-LB for the standard $n_s$, and $\log(10^{10}A_s)$ parameters, as well as the parameters specific to the particle burst model. As expected, for large values of $\log(10^{12}A_I)$ ($\gtrsim 10^{-11})$ we see a decrease in the scalar amplitude $\log(10^{10}A_s)$, but otherwise we do not see much correlation between the primary parameters and those specific to the particle burst model. We see a similar correlation between $n_s$ and $\log(10^{12}A_I)$ and $\log_{10}(k_1)$. Our constraints peak around $\log_{10}(k_1)\simeq -1,$ but all $k_1$ values within our priors are allowed. We see that for larger $k_1$ values, larger $\log(10^{12}A_I)$ is allowed, but as seen in Table~\ref{tab: part_comparison}, the MAP values still prefer a small amplitude even with a large $k_1$. 
\begin{figure}[htbp!]
    \centering
     \begin{subfigure}[t]{0.6\textwidth}
   \includegraphics[width=\textwidth]{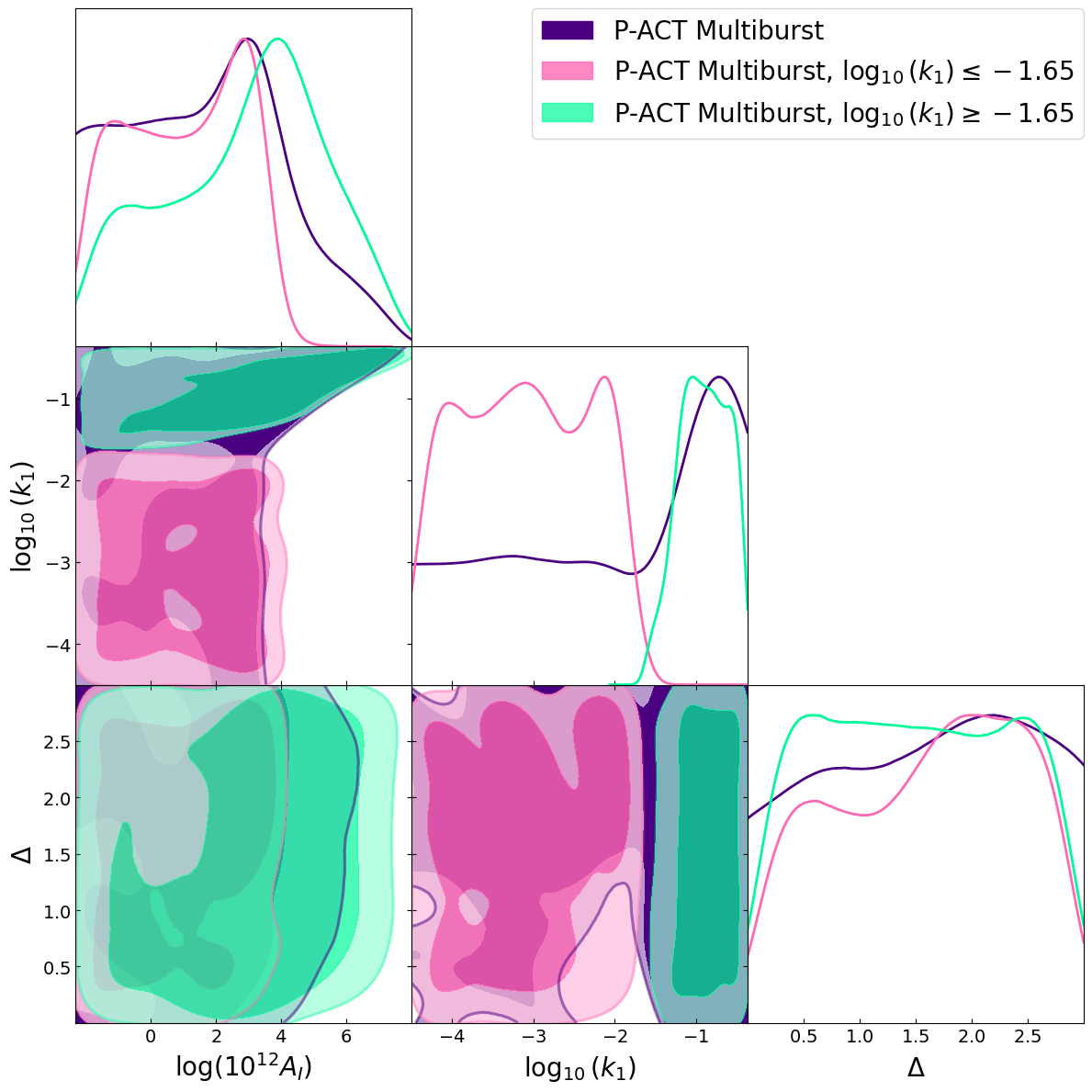}
	\end{subfigure}
    \begin{subfigure}[t]{0.255\textwidth}
   \includegraphics[width=\textwidth]{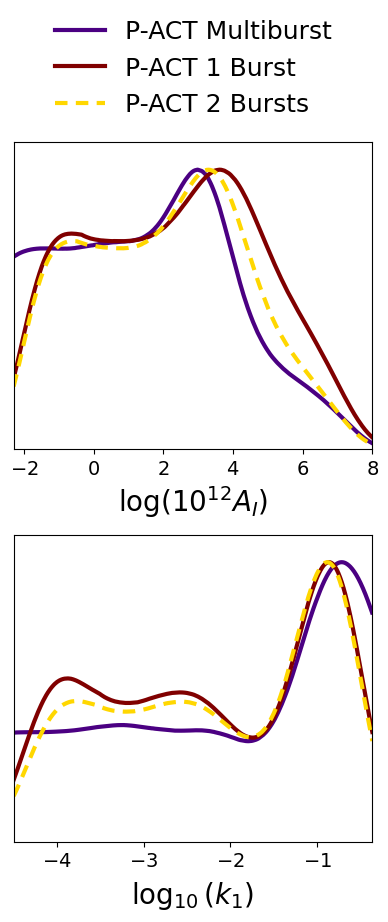}
	\end{subfigure}
    \caption{Showing the changes in parameter constraints for the different $k$-range priors and the number of bursts. \textit{Left:} 1D and 2D posterior distributions using P-ACT for the multiburst particle model for the full $k$-range (indigo), $\log_{10}(k_1) \leq -1.65$ (light green) and $\log_{10}(k_1) \geq -1.65$ (pink). \textit{Right:} 1D posterior distributions using P-ACT for the particle model for 1 burst (maroon), 2 bursts (yellow), and 10 bursts (indigo) for the amplitude and initial burst position.}
     \label{fig: krange and 1b 2b}
\end{figure}
We convert our constraints on $\log(10^{12}A_I)$ to constraints on the dimensionless coupling constant, $g$. These can be seen in Figure~\ref{fig:triangle multi A k Delta} with the MAP values overlaid and the best-fit value ($g = 0.068)$ from Ref.~\cite{Naik2022}. Note that their prior on $\log_{10}(k_1)$ is truncated at $-1.0$ compared to ours, which extends to $\log_{10}(k_1)=-0.365$ due to the inclusion of our high-$\ell$ ACT data. In addition, they do most of their analysis assuming $0 \leq \Delta \leq 1$, as they also study the single burst model extensively. As we focus primarily on the multiburst model, we consider a large range of $0 \leq \Delta \leq 3$. Despite our increased parameter space, we obtain MAP values of $g = 0.040\, (0.034)$ for P-ACT (P-ACT-LB). Physically, tightening the constraints $g$ limits the interaction strength between the inflaton and iso-inflaton, which in turn limits the number density of the iso-inflaton. Additionally, because the time-dependent mass of the iso-inflaton is proportional to $g$ \citep{1702.07661.2017JCAP...05..054P}, restricting the coupling constant constraint also tightens the bounds on the mass of the iso-inflaton. 

To investigate the scale preference, we split the prior range for $\log_{10}(k_1)$ to $\log_{10}(k_1)\geq -1.65$ and $\log_{10}(k_1)\leq-1.65$ regions, as shown in the left panel of Figure~\ref{fig: krange and 1b 2b}. This value of $\log_{10}(k_1)$ was chosen as it corresponds to where the 1D posterior for $\log_{10}(k_1)$  started increasing away (for $\log_{10}(k_1) \sim -1$) from the fairly flat posterior value on larger scales. As expected, larger amplitude values are favoured for $\log_{10}(k_1) \geq -1.65$ and smaller values for $\log_{10}(k_1) \leq -1.65$. Despite this, the MAP values show an opposite trend, as can be seen in Table~\ref{tab: part_comparison}. This is likely due to the MAP value $\Delta = 0.629$ which in turn allows for a larger amplitude range, as can be seen in the left panel of Figure~\ref{fig: krange and 1b 2b}, where there is a pocket of low probability density for both large $\Delta$ and low amplitude.

\begin{figure}
\centering
\includegraphics[width=0.8\linewidth]{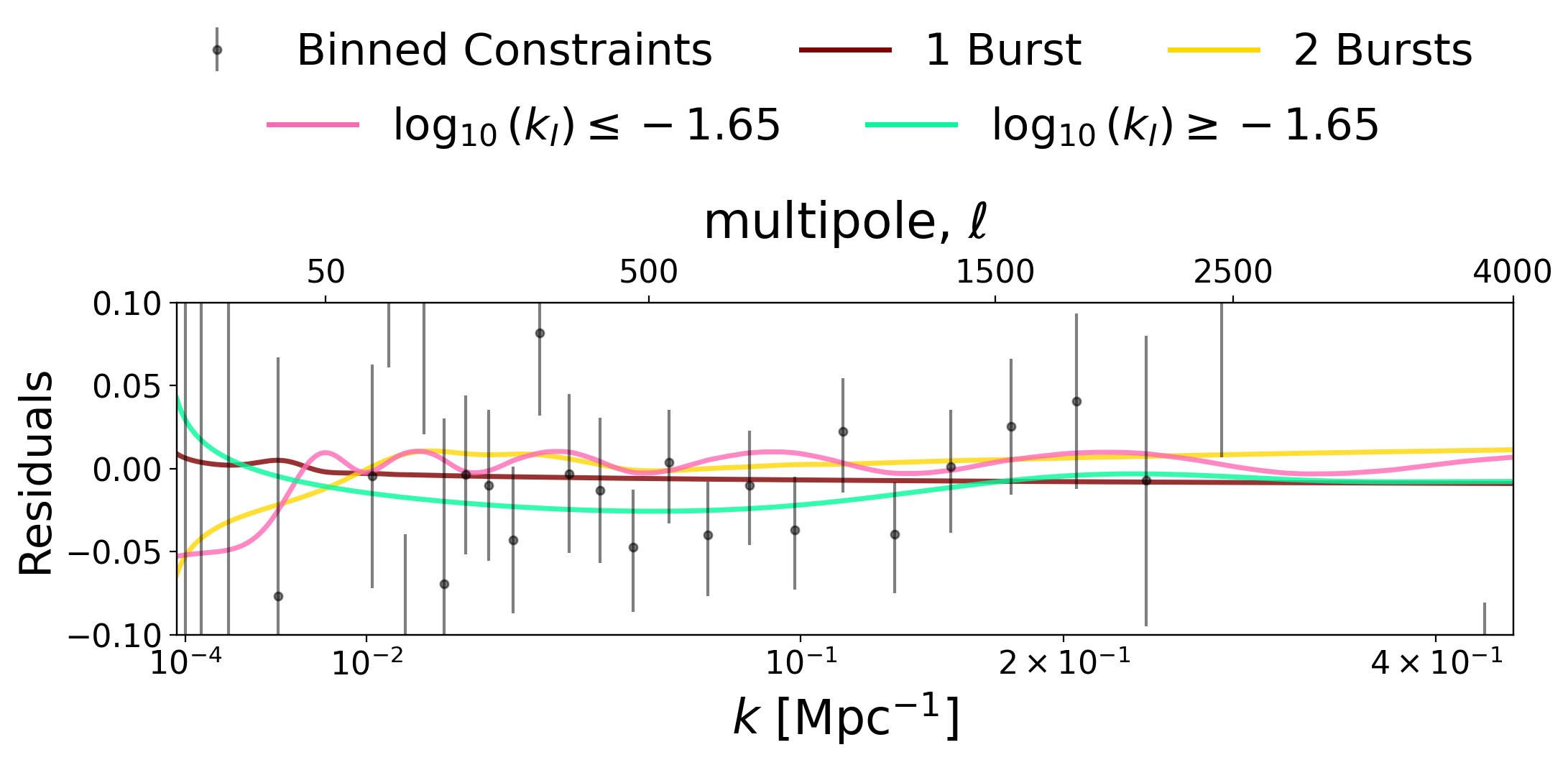}
\caption{Comparing the MAP spectra for the 1 burst, 2 bursts, and different $k$-range particle burst priors. Our binned $P_{\mathcal{R}}(k)$ constraints from \cite{ACTextended2025JCAP...11..063C} are shown as error bars. The residuals are with respect to the P-ACT $\Lambda$CDM best fit. Note that  the $x$-axis is scaled as $k^{0.5}$ in order to best show the small scales.} \label{fig: all ext part model binned}
\end{figure}

Next, we explore the effects of the number of particle bursts allowed in our model, using the P-ACT dataset, focusing specifically on 1, 2, and 10 bursts. As shown in Figure~\ref{fig: multi 1b 2b full} in Appendix Section~\ref{ssec: full part}, most parameters do not change significantly depending on the number of bursts included. We do, however, observe moderate changes for $\log(10^{12}A_I)$ and $\log_{10}(k_1)$ which can be seen in the right panel of Figure~\ref{fig: krange and 1b 2b}. For 1 or 2 bursts of particle production, we see a shift to slightly smaller $k_1$ values, in particular for 1 burst, which is reflected in the MAP values in Table~\ref{tab: osc_comparison}. Additionally, we see a shift to large amplitude values, again in particular for 1 burst. In contrast, we do obtain a larger MAP value for the amplitude for 2 bursts compared to 1. These MAP values fit the data well, as demonstrated in Figure~\ref{fig: all ext part model binned}, which compares the MAP spectra for 1 burst, 2 bursts, and the different $k$-range priors to the binned PPS constraints. 

In Ref.~\citep{Hidde}, they find a hint for 1 burst of particle production for $ 0.3 \lesssim k \lesssim 1 \: \text{Mpc}^{-1}$. In our analysis, we see a peak at $\log_{10}(k_1) \simeq -1$, but the posterior remains broad, leaving all higher values supported up to our prior boundary of $\log_{10}(k_1) = -0.365$ ($k_1 = 0.43 \: \text{Mpc}^{-1}$). Given that we consider different models for inflationary particle production and that they conduct a grid search for the position of the feature and we allow the position to vary with the other parameters, these results are broadly compatible.

\begin{table*}[htbp!]
\centering
\begin{adjustbox}{max width=\textwidth}
\begin{tabular}{l|cc|c|c|c|c}
\toprule
 & \multicolumn{2}{c|}{\textbf{Multiburst}} & & & & \\
\cline{2-3}
\textbf{Parameter} & \textbf{P-ACT} & \textbf{P-ACT-LB} & \textbf{1 Burst} & \textbf{2 Bursts} & $\mathbf{\log_{10}(k_I) \geq -1.65}$ & $\mathbf{\log_{10}(k_I) \leq -1.65}$ \\
\midrule
$\log(10^{10} A_\mathrm{s})$ & 3.05 & 3.05 & 3.05 & 3.05 & 3.04 & 3.03 \\ 
$n_\mathrm{s}$ & 0.971 & 0.973 & 0.970 & 0.974 & 0.967 & 0.970 \\
$\log(10^{12} A_I)$ & 2.10 & 1.56 & 1.80 & 2.70 & 3.04 &  3.80\\ 
$\log_{10}(k_1)$ & -1.08 & -1.29 & -3.00 & -2.37 & -1.23 & -2.77 \\
$\Delta$ & 0.890 & 0.785 & -- & 0.737 & 0.629 & 0.876 \\
\midrule 
$\Delta\chi^2$ & $-1.35$ & $-0.12$ & $-0.50$ & $-2.90$ & $-3.55$ & $-1.40$ \\
$\ln B$ & $-0.08$ & $-0.23$ & $0.03$ & $-0.05$ & $0.30$ & $-0.40$ \\
\bottomrule
\end{tabular}
\end{adjustbox}
\caption{Combined MAP values for the particle burst model from the \texttt{Cobaya} minimizer, including the P-ACT-LB dataset. $\Delta\chi^2$ and $\ln B$ are relative to $\Lambda$CDM. The $\Delta\chi^2$ values are computed using the MAP values, whereas the $\ln B$ values are computed using the evidence value from \texttt{pocoMC}. Negative $\Delta\chi^2$ indicates an improved fit and negative $\ln B$ indicates a preference for $\Lambda$CDM.}
\label{tab: part_comparison}
\end{table*}

\subsection{Model comparison}\label{ssec: model comparison}
Previous subsections presented specific constraints on individual models, however we must evaluate the preference for any model (whether of the general oscillation or particle burst form) to the base $\Lambda$CDM cosmology. We evaluate the models using both the $\Delta\chi^2$ values for the MAP fits as well as the Bayesian evidence, $\ln B$, computed using \texttt{pocoMC}. Negative $\Delta\chi^2$ indicates an improved fit and negative $\ln B$ indicates a preference for $\Lambda$CDM. Evidence strength follows the Jeffreys’ scale: Inconclusive ($|\ln B| < 1$), Weak (1 to 2.5), Moderate (2.5 to 5), Strong ($>5$) \citep{2008ConPh..49...71T}.

For all the general oscillating models, we observed an improved fit across all dataset combinations with $\Delta\chi^2$ values ranging from $-3.55$ to $-8.93$. However, despite these improved fits, the Bayesian evidence all consistently have a moderate preference for $\Lambda$CDM. The $\ln B$ values range from $-2.60$ to $-3.84$ which indicates that the added parameters and complexity of these oscillatory models are not statistically justified with current data.

The particle production model for the different priors and number of bursts also has improved fits compared to $\Lambda$CDM, with $\Delta\chi^2$ values ranging from $-0.12$ to $-3.55$. But the Bayesian evidence values are all inconclusive. Most have a negative $\ln B$ indicating a slight inconclusive preference for $\Lambda$CDM ranging from $-0.05$ to $-0.40$. But, the evidence is positive but inconclusive for the single burst of particle production ($\ln B = 0.03$) and for the $k_1$ prior of $\log_{10}(k_1) \geq -1.65$ ($\ln B = 0.30$). This $k_1$ prior also has the most significant $\Delta\chi^2$ value, of $-3.55$.

\begin{figure}
    \centering
    \begin{subfigure}[t]{0.88\linewidth}
   \includegraphics[width=\textwidth]{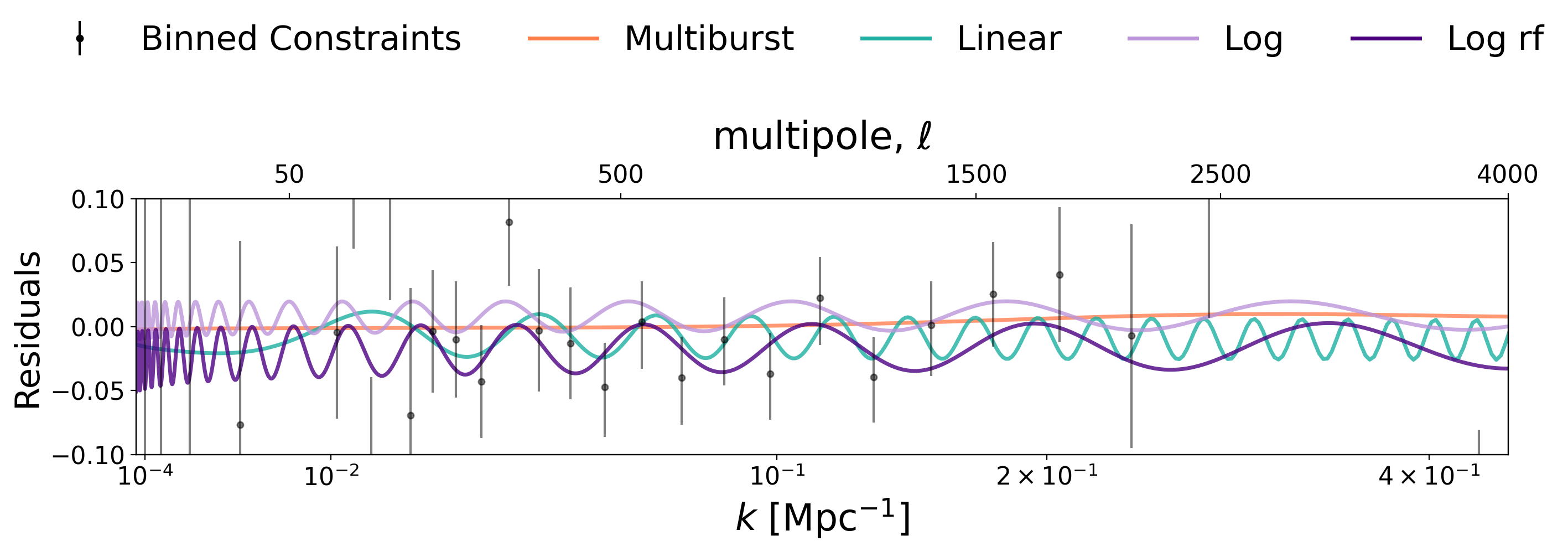}
	\end{subfigure}
    \begin{subfigure}[t]{0.9\linewidth}
   \includegraphics[width=\textwidth]{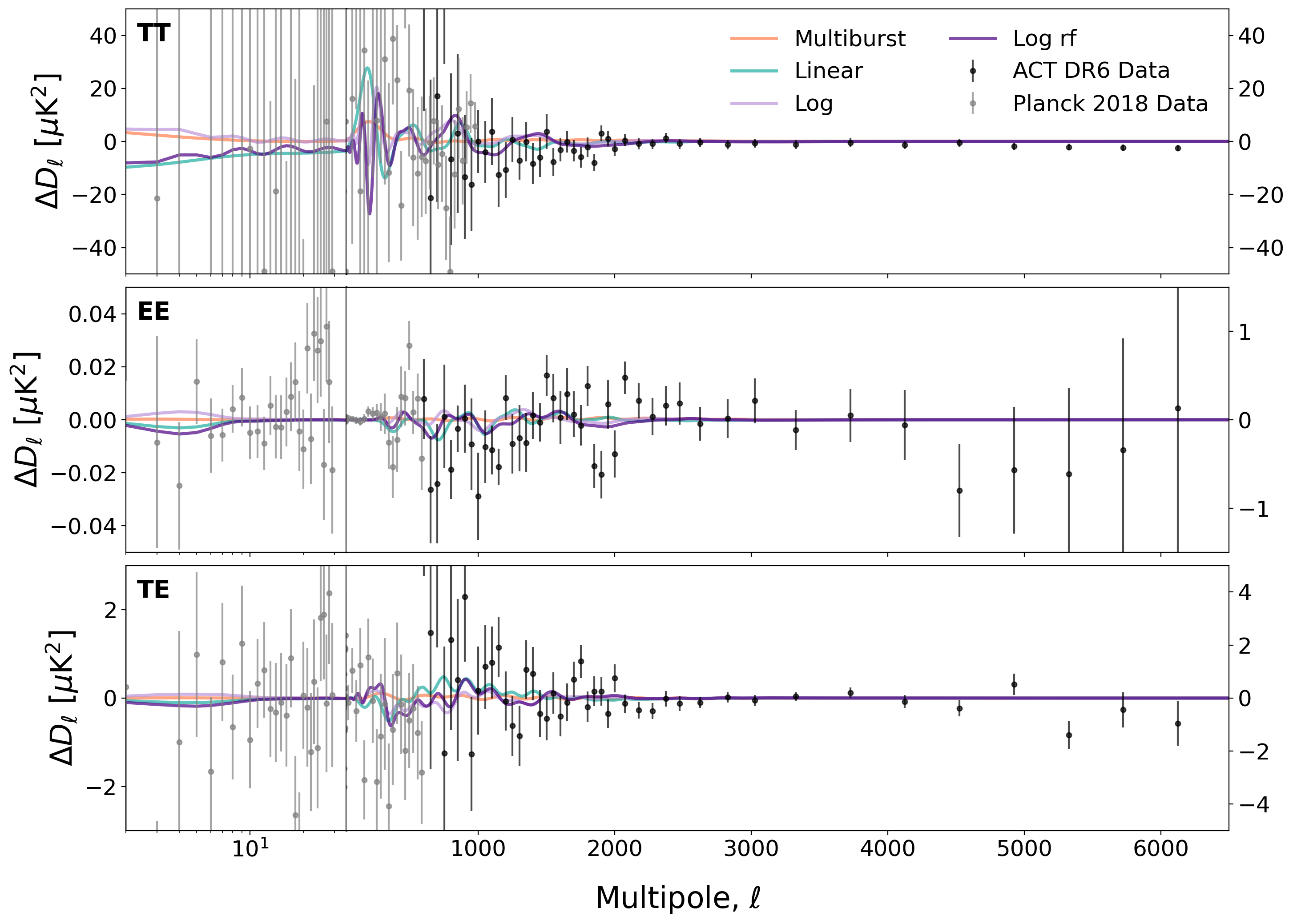}
	\end{subfigure}
    \caption{Comparing the MAP spectra for the different oscillating $P_{\mathcal{R}}(k)$ models. \textit{Top:} Our binned $P_{\mathcal{R}}(k)$ constraints from \cite{ACTextended2025JCAP...11..063C} are shown as error bars. Note that  the $x$-axis is scaled as $k^{0.5}$ in order to best show the small scales. \textit{Bottom:} Comparing the resulting $D_{\ell}$ for the MAP spectra with the \textit{Planck} 2018 and ACT DR6 data. The residuals are with respect to the P-ACT $\Lambda$CDM best fit.}
     \label{fig: all models}
\end{figure}

We present the P-ACT MAP spectra for all models and compare them to the binned $P_{\mathcal{R}}(k)$ results in top panel of Figure~\ref{fig: all models}. As can be seen, due to the particle burst model preferring both a low amplitude and large $k_1$ value, the resulting spectrum does not have any visible oscillations and is very close to $\Lambda$CDM. Conversely, the general oscillation models \textit{cannot shift the initial wavenumber} where the oscillations occur, and thus all attempt to fit the fluctuations in the data. This can be seen further in the bottom panel of Figure~\ref{fig: all models}, which shows the residuals of these MAP spectra and the P-ACT $\Lambda$CDM best fit as well as the \textit{Planck} 2018 and ACT DR6 data.

\section{Conclusions} \label{sec: conclusions}

In this paper, we constrained general oscillatory PPS models as well as oscillations induced by inflationary particle production using data from \textit{Planck}, ACT DR6, and DESI DR2. In order to model the non-linear matter power spectrum with our accuracy settings and highly oscillatory PPS, we implemented the corrections outlined in Ref. \cite{WMP2019PhRvR...1c3209B} in \texttt{CAMB}. For the general oscillating models, we found 95\% CL upper limits on the oscillation amplitudes of $A_{\text{lin}} < 0.021$, $A_{\text{log}} < 0.022$, and $A_{\text{log rf}} < 0.023$, which restricts the amplitude to $\sim 2\%$ of the $A_s$ value. While these general oscillating models all had improved fits compared to $\Lambda$CDM with $\Delta\chi^2$ values ranging from $-3.55$ to $-8.93$, the Bayesian evidence computed with \texttt{pocoMC} consistently showed a moderate preference for $\Lambda$CDM. The $\ln B$ values ranged from $-2.60$ to $-3.84$, indicating that the added complexity of these models is not justified with current data. Additionally, we observe a phase shift between \textit{Planck} and P-ACT similar to our previous binned PPS reconstruction.

We investigated inflationary particle production for 1, 2, and 10 burst models, utilizing a Taylor expansion to alleviate numerical instabilities present at small $k/k_i$. We derived constraints on the coupling constant from our amplitude constraints, obtaining MAP values of the dimensionless coupling constant $g = 0.040$ for P-ACT and $g = 0.034$ for P-ACT-LB. Similar to the general oscillations, the particle burst models had an improved fit compared to $\Lambda$CDM, with $\Delta\chi^2$ values ranging from $-0.12$ to $-3.55$, but the Bayesian evidence was inconclusive for all models and prior ranges considered. Due to the multiburst model preferring both a low amplitude and large $k_1$, the resulting PPS lacks visible oscillations and remains very close to $\Lambda$CDM. We do note that these results consider a constant spacing between bursts, i.e. for constant $\Delta$. As we showed in Figure~\ref{fig: delta shift}, this assumption results in incorrect estimates for the $k_i$ values where the bursts occur of up to 51.4\% and 117\% for the monomial and $\alpha$-attractor potentials, respectively. We leave the correction of this effect for future work. 

In summary, these results indicate that while current data permits marginal improvements in fit with oscillatory PPS, the Bayesian evidence reinforces the statistical preference for the concordance $\Lambda$CDM model. Upcoming surveys, such as the Simons Observatory \citep{ade/etal:2019, so_collaboration:2025}, will increase our observational precision even more on both large and small scales, which will allow for unprecedented constraints on non-standard inflationary models. Further work addressing the cumulative effects of burst spacing, especially for the small-scale inflationary potentials consistent with current data, is crucial to accurately model inflationary particle production. Furthermore, the integration of samplers such as \texttt{pocoMC} is essential for sampling these complex, multimodal parameter spaces as we continue to explore the inflationary landscape.

\acknowledgments
We thank Adam Hincks for comments and discussions on the paper and analysis. SKN acknowledges support from the Queen Elizabeth II Graduate Scholarship in Science and Technology, NSERC Canada Graduate Scholarships, Walter C. Sumner Memorial Fellowship, and the Lachlan Gilchrist Fellowship. RH acknowledges support from the NSERC Discovery Grant Program RGPIN-2018-05750 and the Connaught Fund. The Dunlap Institute is funded through an endowment established by the David Dunlap family and the University of Toronto. The authors at the University of Toronto acknowledge that the land on which the University of Toronto is built is the traditional territory of the Wendat Nation, the Seneca, and the Mississaugas of the Credit. Today, this meeting place is still the home to many Indigenous people from across Turtle Island, and they are grateful to have the opportunity to work on this land.


\providecommand{\href}[2]{#2}\begingroup\raggedright\endgroup

\appendix

\section{Binned PPS updates with DESI DR2}\label{sec: binned}

We update our previous binned PPS reconstruction from C25 to use the DESI DR2 BAO measurements. We constrain $e^{-2\tau}P_{\mathcal{R}}(k_{i})$ (in order to account for the ${A_s-\tau}$ degeneracy) for wavenumber bins centered at $k = 10^{-4}, 10^{-3.5}, 10^{-3}, 10^{-2.5}$ and then 26 equally-spaced logarithmic bins from ${0.011\lesssim k ~/\mathrm{Mpc^{-1}}\lesssim 0.43}$, where $k_{i+1} \simeq 1.16k_i$. For this analysis, we use the traditional MCMC sampler within \texttt{Cobaya} to remain consistent with our previous work in C25. The priors used for the standard cosmological parameters and for each wavenumber bin on our reconstruction are outlined in Table~\ref{tab:priors binned}. The results can be seen in Figure \ref{fig:binned Pk} alongside the results from C25. We do not see a large difference between our P-ACT-LB and P-ACT-LB$_{\text{DR}2}$ results, but we do note that there are marginal shifts in the central values.

\begin{table}
\centering
\small
\begin{tabular}{l|c|l|c}
\toprule
\textbf{Parameter} & \textbf{Prior} & \textbf{Parameter} & \textbf{Prior} \\ 
\midrule
\multicolumn{4}{l}{\textbf{$\mathbf{\Lambda}$CDM}} \\ 
\midrule
$\Omega_\mathrm{b}h^2$ & $[0.017, 0.027]$ & $\log(10^{10} A_\mathrm{s})$ & $[2.9, 3.2]$ \\
$\Omega_\mathrm{c}h^2$ & $[0.09, 0.15]$ & $n_\mathrm{s}$ & $[0.92, 1.02]$ \\
$\theta_\mathrm{MC}$ & $[0.01038, 0.01044]$ & $\tau_\mathrm{reio}$ & $[0., 0.1]$ \\ 
\midrule
\multicolumn{4}{l}{\textbf{ACT Specific}} \\ 
\midrule
$A_{\rm ACT}$ & \multicolumn{3}{l}{normal, loc: 1.0, scale: 0.003} \\
$p_{\rm ACT}$ & \multicolumn{3}{l}{$[0.5, 1.5]$} \\ 
\midrule
\multicolumn{4}{c}{\textbf{Binned Priors}} \\ 
\midrule
$k \times 10^{2}$ [Mpc$^{-1}$] & Range $\times 10^{9}$ & $k \times 10^{2}$ [Mpc$^{-1}$] & Range $\times 10^{9}$ \\ 
\midrule
0.01 & 0.00 -- 15.0 & 5.42 & 1.50 -- 2.10 \\
0.03 & 0.10 -- 10.0 & 6.29 & 1.80 -- 2.05 \\
0.10 & 0.30 -- 9.00 & 7.29 & 1.70 -- 2.00 \\
0.32 & 0.50 -- 8.00 & 8.45 & 1.75 -- 2.00 \\
1.06 & 1.40 -- 2.60 & 9.81 & 1.76 -- 1.95 \\
1.23 & 1.40 -- 2.40 & 11.4 & 1.70 -- 2.00 \\
1.43 & 1.40 -- 2.40 & 13.2 & 1.70 -- 1.93 \\
1.66 & 1.60 -- 3.30 & 15.3 & 1.65 -- 2.00 \\
1.92 & 1.60 -- 2.30 & 17.7 & 1.51 -- 2.12 \\
2.23 & 1.60 -- 2.20 & 20.6 & 1.32 -- 2.30 \\
2.59 & 1.60 -- 2.20 & 23.8 & 1.04 -- 2.55 \\
3.00 & 1.60 -- 2.20 & 27.7 & 0 -- 4.26 \\
3.48 & 1.68 -- 2.14 & 32.1 & 0 -- 10.7 \\
4.03 & 1.75 -- 2.06 & 37.2 & 0 -- 20.0 \\
4.68 & 1.50 -- 2.10 & 43.1 & 0 -- 8.46 \\ 
\bottomrule
\end{tabular}
\caption{Priors for all parameters used in the binned PPS reconstruction. The priors are flat unless otherwise specified.}
\label{tab:priors binned}
\end{table}

\section{Numerical instability of the particle burst model}\label{sec: taylor}

\begin{figure}
    \centering
     \begin{subfigure}[t]{1\textwidth}
   \includegraphics[width=\textwidth]{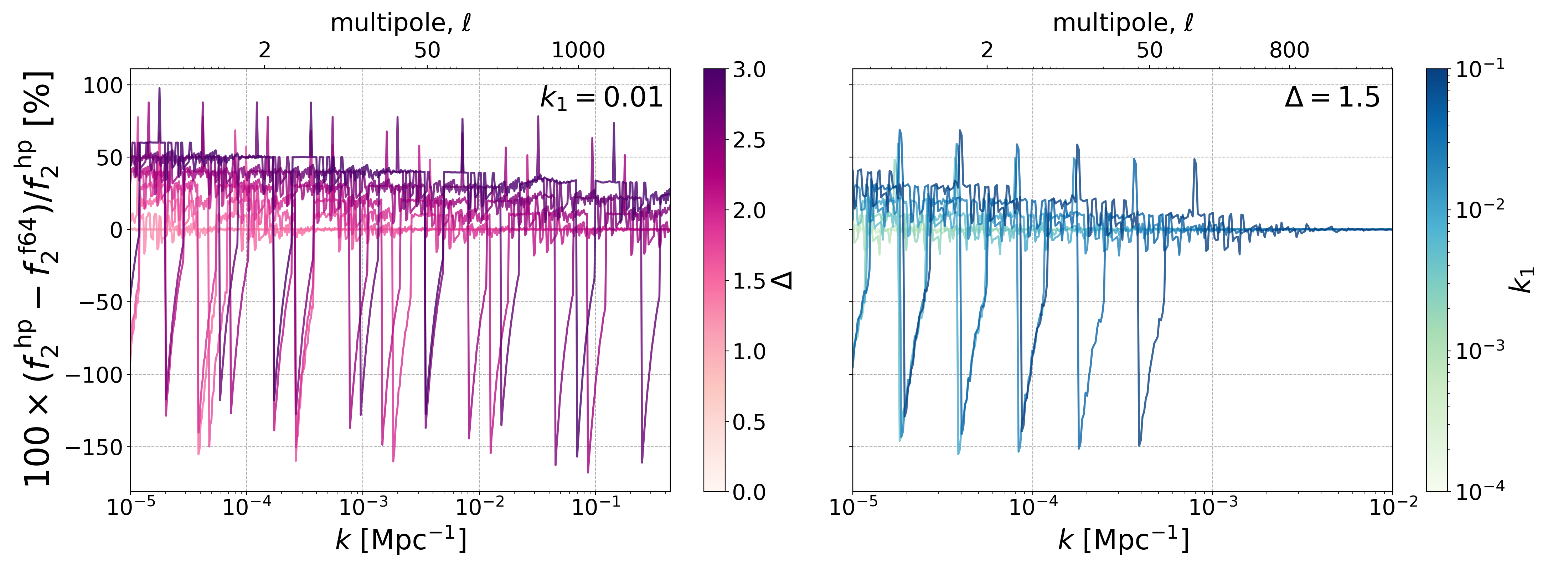}
	\end{subfigure}
    \begin{subfigure}[t]{0.45\textwidth}
   \includegraphics[width=\textwidth]{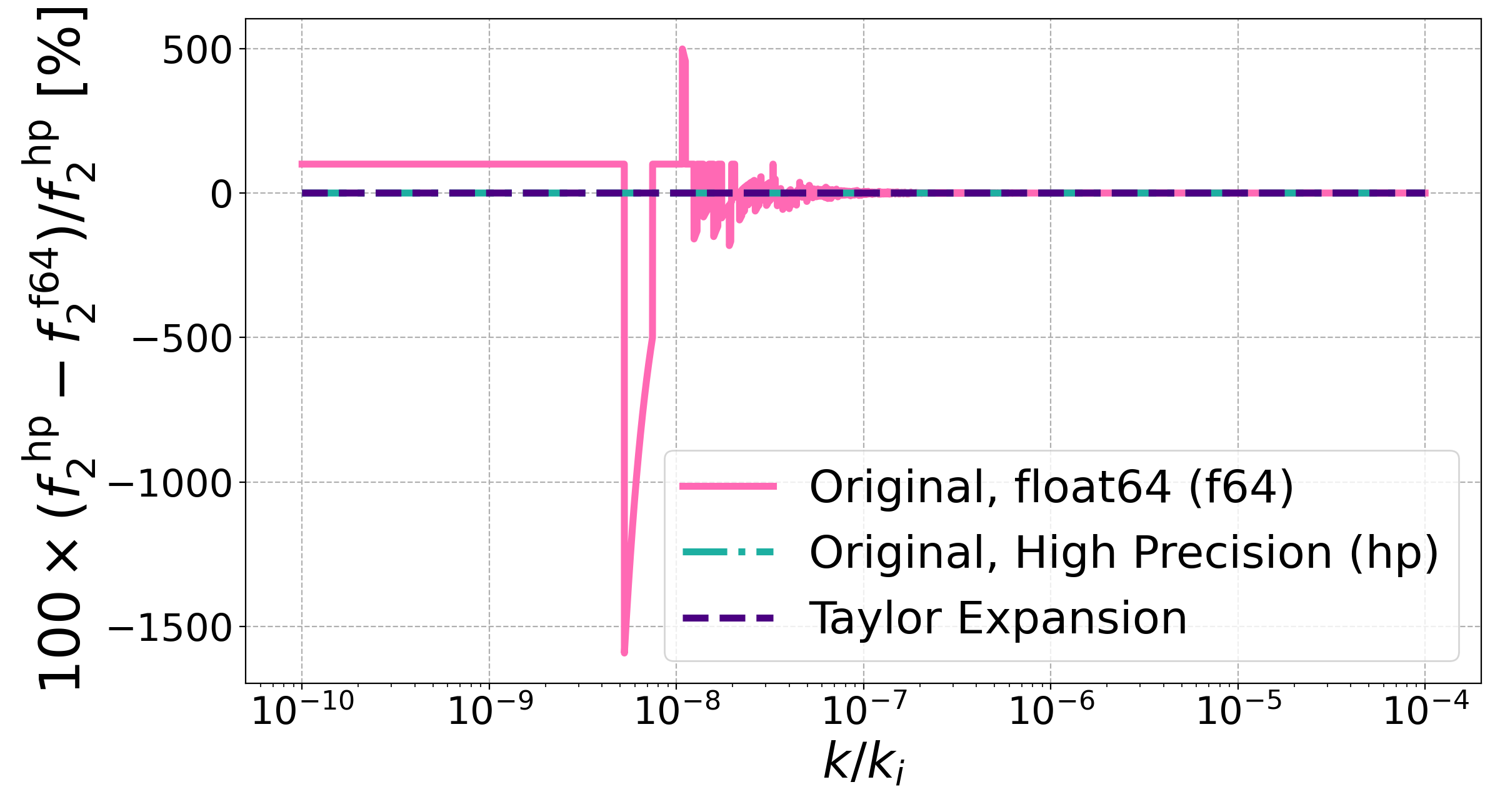}
	\end{subfigure}
    \caption{Fractional residuals comparing $f_2$ from Eq.~\ref{eq: f2} for 50 significant digits (hp) and standard float64 (f64) machine precision ($\sim 15$ significant digits). Note that the top two panels are shown as functions of $k$, but $k/k_i$ is input to $f_2$ as normal, whereas the bottom panel is shown as a function of $k/k_i$ to illustrate why the transition point to use a Taylor expansion is $k/k_i < 10^{-5}$.}
     \label{fig: taylor}
\end{figure}

The particle production model given in Eq.~\ref{eq:powerspec multi} exhibits numerical instabilities at small values of $x_i \equiv k/k_i$, originating from the $f_2(x_i)$ term. This can be seen in the top panel of Figure~\ref{fig: taylor}, where we show the normalized residuals for the standard 64-bit floating-point (float64, f64) machine precision and 50-digit arbitrary precision (hp). We show the $f_2(x_i)$ term for a variety of $\Delta$ values with $k_1 = 0.01$ on the left and a variety of $k_1$ values for $\Delta = 1.5$ on the right. We see up to a 150\% difference and a spike pattern due to these numerical instabilities. In order to resolve this, implement a Taylor expansion for the $f_2(x_i)$ term for $x_i = k/k_i < 10^{-5}$. We chose this value to ensure we have a buffer before the numerical issues become present at $k/k_i \simeq 10^{-7}$ as can be seen in the bottom panel of Figure~\ref{fig: taylor}.

\section{Settings and \texttt{CAMB} modifications}

\subsection{Settings} \label{ssec: settings}

To ensure accurate computation in \texttt{CAMB}, we use the accuracy settings outlined in Appendix A of C25. To sample our multimodal parameter space, we use the \texttt{pocoMC} with 4096 active particles (\texttt{n$_{\text{active}}:$ 4096}) and a total of 8192 effectively independent samples (\texttt{n$_{\text{total}}:$ 8192}). Similarly, we use 8192 importance samples to estimate the Bayesian evidence (\texttt{n$_{\text{evidence}}:$ 8192}). We use the t-preconditioned Crank-Nicolson sampler (\texttt{sample: tpcn}) and we precondition with a Neural Spline Flow with 6 transformations (\texttt{flow: nsf6}). The other parameters remain at their default values.

\subsection{Non-linear matter power spectrum modifications} \label{ssec: matter power}

\begin{figure}
\centering
\includegraphics[width=0.7\linewidth]{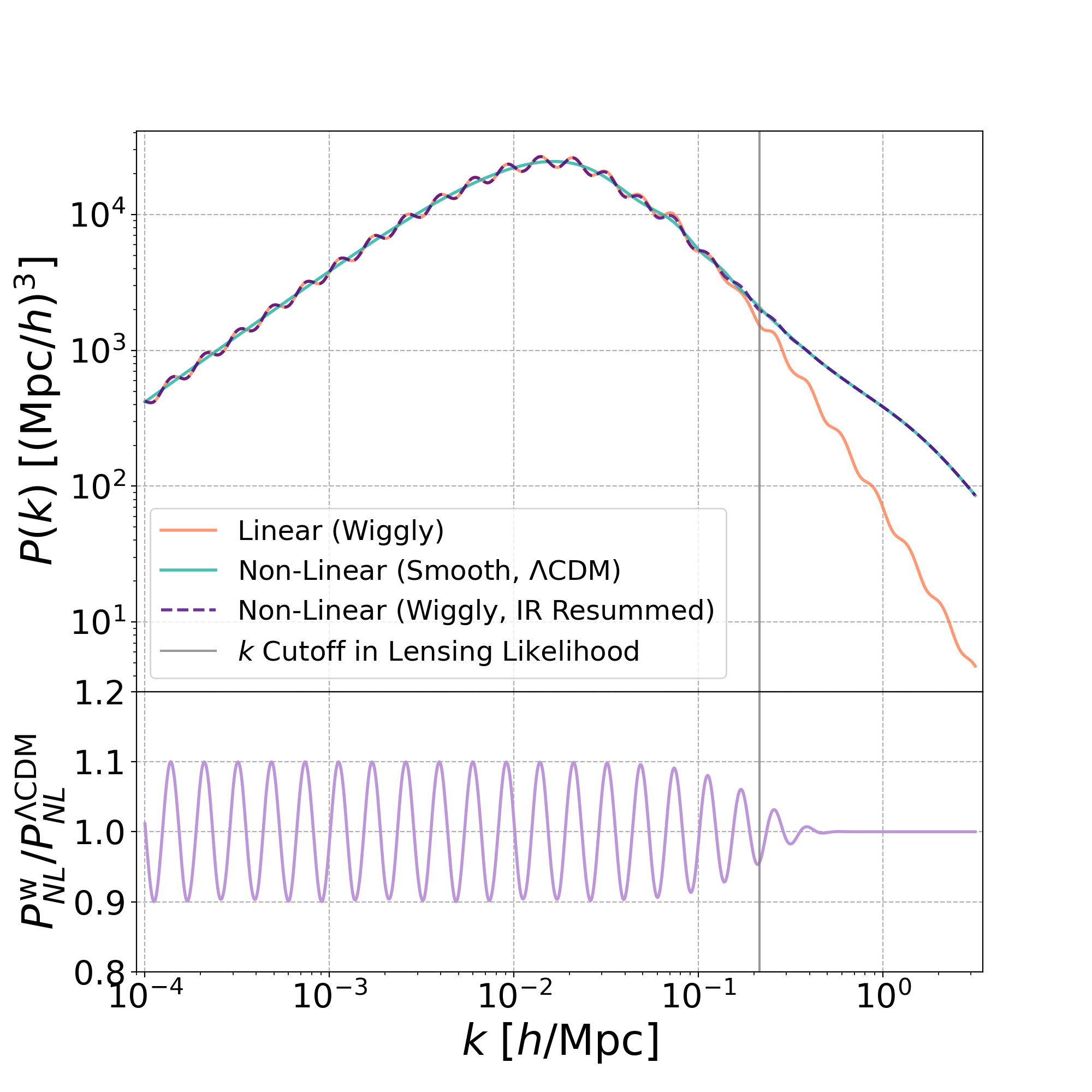}
\caption{Comparison of the linear and non-linear matter power spectra computed using the accuracy settings for \texttt{CAMB} shown described in Appendix A of C25. We use the IR resummation corrections to the non-linear power spectrum outlined in Eq.~\ref{eq: nonlinear matter power} and Ref.~\cite{WMP2019PhRvR...1c3209B}. The bottom panel displays the ratio of oscillatory and baseline $\Lambda$CDM power spectra.} \label{fig: nl matterpower}
\end{figure}

Following Ref.~\citep{WMP2019PhRvR...1c3209B}, we modify the non-linear matter power spectrum computations in \texttt{HMCode} within \texttt{CAMB}. For the linear and logarithmic models discussed in this paper, Ref.~\citep{WMP2019PhRvR...1c3209B} shows that the matter power spectrum can be approximated as,

\begin{equation}\label{eq: nonlinear matter power}
P_m(k) \approx P_m^{\text{nw}}(k) + e^{-k^2\Sigma^2_{\text{BAO}}}\Big[P_{\text{BAO}}^{\text{nw}}(k) + P^{\text{w}}(k)\Big],    
\end{equation}

\noindent where $P_m(k)$ is the matter power spectrum, $\Sigma_{\text{BAO}}$ is the BAO damping scale, $P_{\text{BAO}}^{\text{nw}}(k)$ is the BAO power spectrum, and $P^{\text{w}}(k)$ is the oscillatory component of the PPS. Figure~\ref{fig: nl matterpower} shows the $\Lambda$CDM non-linear matter power spectrum as well as the linear and non-linear matter power spectra for the logarithmic spacing oscillation model. We indicate the $k$ cutoff in the lensing likelihood.

\section{Full results} \label{sec: full results}

\subsection{General oscillations} \label{ssec: full osc}

\begin{figure}
    \centering
    \includegraphics[width=\linewidth]{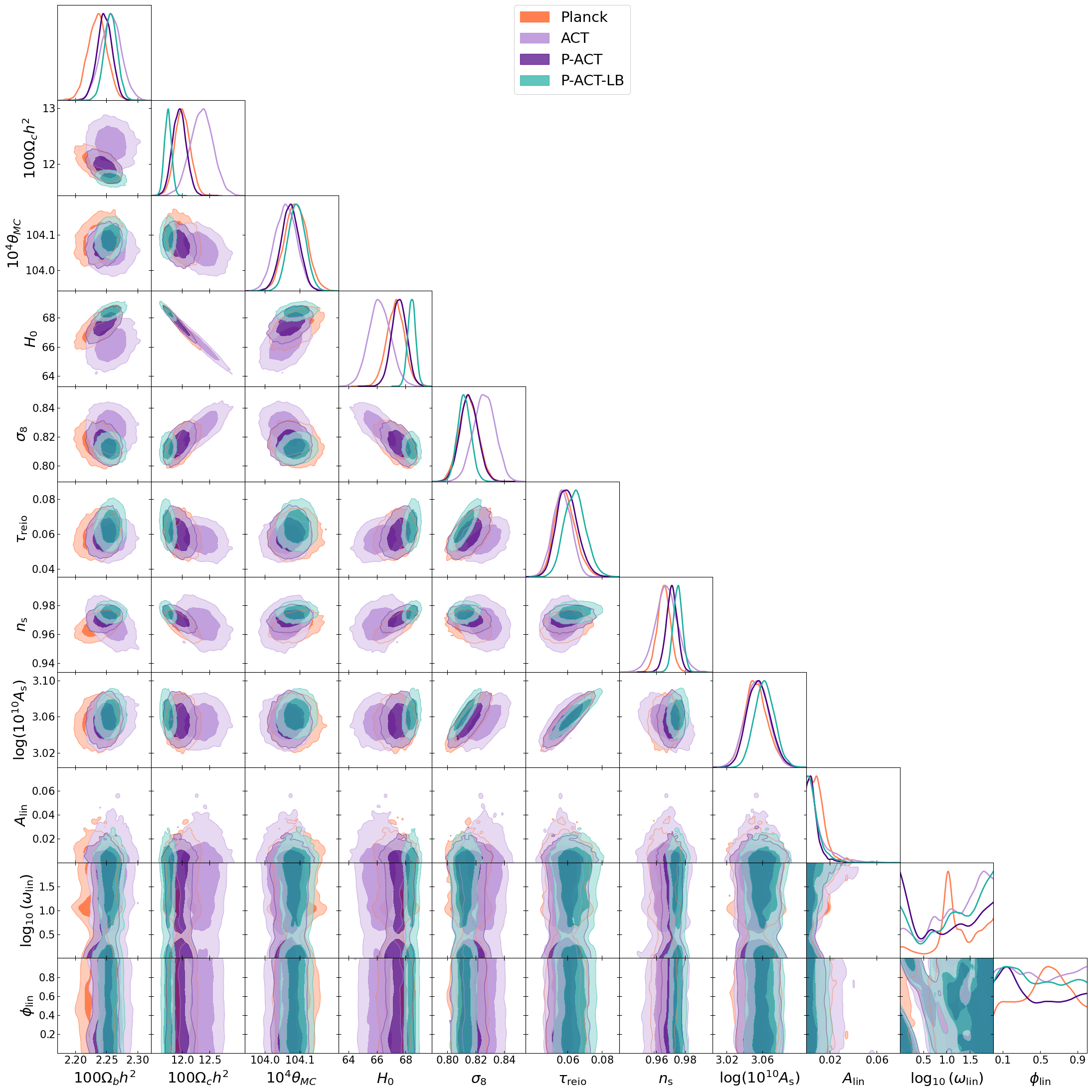}
    \caption{Marginalized 1D and 2D posterior distributions for the linearly spaced oscillating $P_{\mathcal{R}}(k)$ model outlined in Eq.~\ref{eq: oscillating power spec}. We show both the standard $\Lambda$CDM parameters and the oscillation parameters. We show the results for \textit{Planck} (coral), ACT (light purple), P-ACT (indigo), and P-ACT-LB (teal).}
    \label{fig: lin all}
\end{figure}

\begin{figure}
    \centering
    \includegraphics[width=\linewidth]{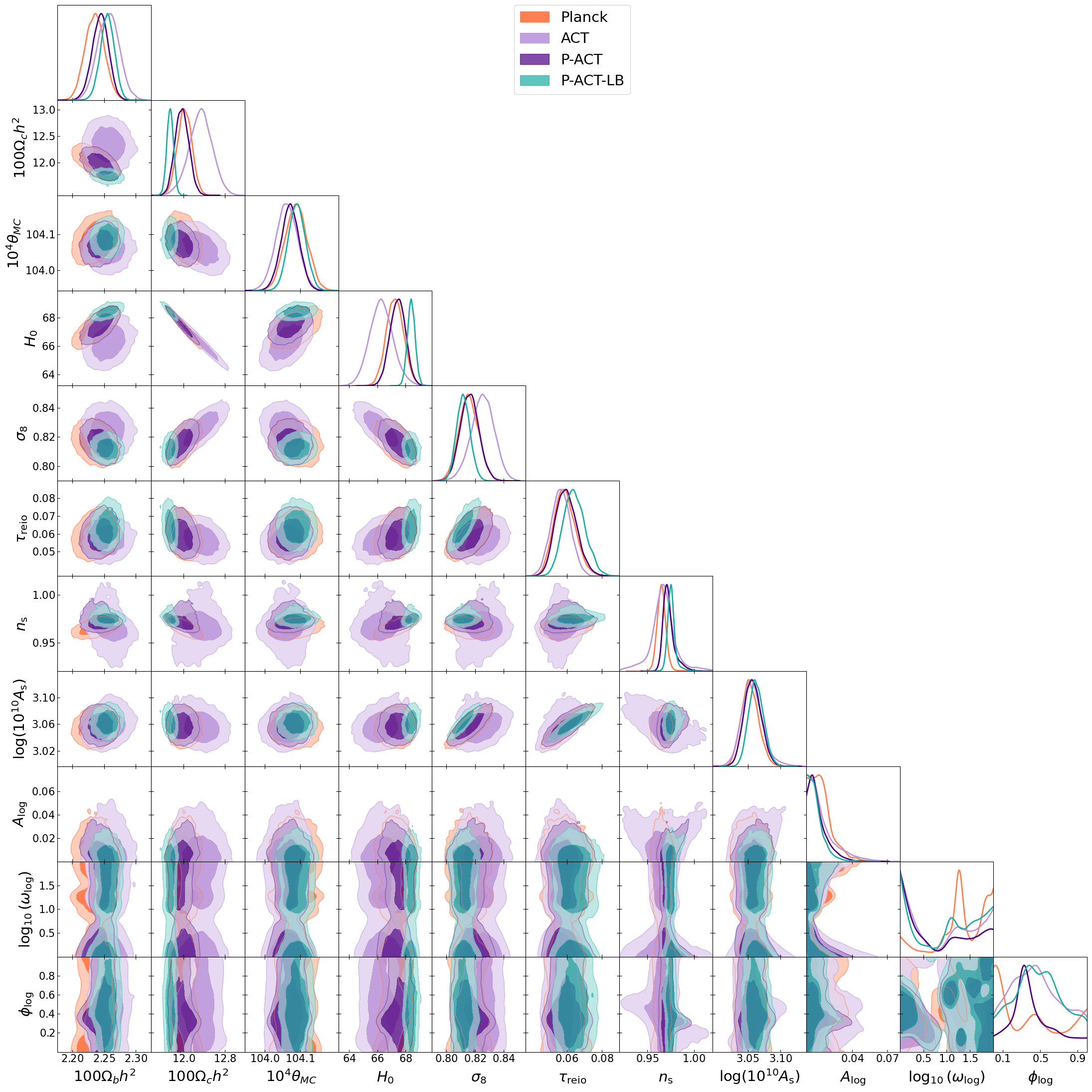}
    \caption{Same as Figure~\ref{fig: lin all} but for logarithmically spaced oscillations.}
    \label{fig: log all}
\end{figure}

\begin{figure}
    \centering
    \includegraphics[width=\linewidth]{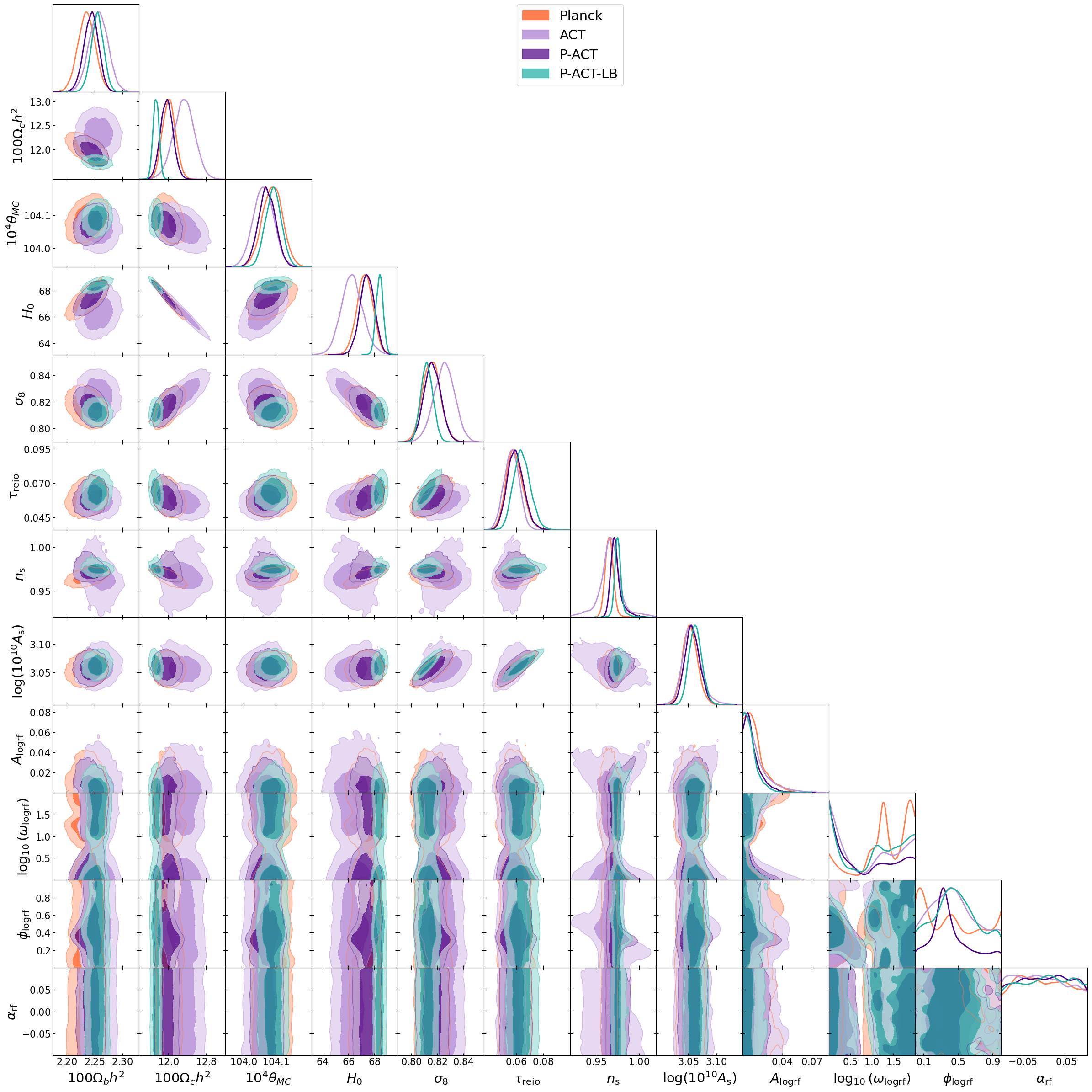}
    \caption{Same as Figure~\ref{fig: lin all} but for logarithmically spaced oscillations with a running frequency.}
    \label{fig: logrf all}
\end{figure}

We provide the triangle plots for the general oscillating models discussed in Section~\ref{ssec: osc models}, for both the $\Lambda$CDM and oscillating parameters. Figures~\ref{fig: lin all}, \ref{fig: log all}, and \ref{fig: logrf all} show the 1D and 2D marginalized posterior distributions for the linearly, logarithmically, and logarithmically with a running frequency spaced models, respectively. In each figure, we compare the constraints derived from \textit{Planck} alone, ACT alone, P-ACT, and P-ACT-LB. In general, we do not see a large change in the $\Lambda$CDM parameters when comparing the $\Lambda$CDM-only runs, except for $A_s$ and $n_s$ as expected. Additionally, our results for $\alpha_{\text{rf}}$ are unconstrained. Thus, the constraints are very similar for both the logarithmic and logarithmic with a running frequency.

\subsection{Particle burst} \label{ssec: full part}

\begin{figure}
    \centering
    \includegraphics[width=1.0\linewidth]{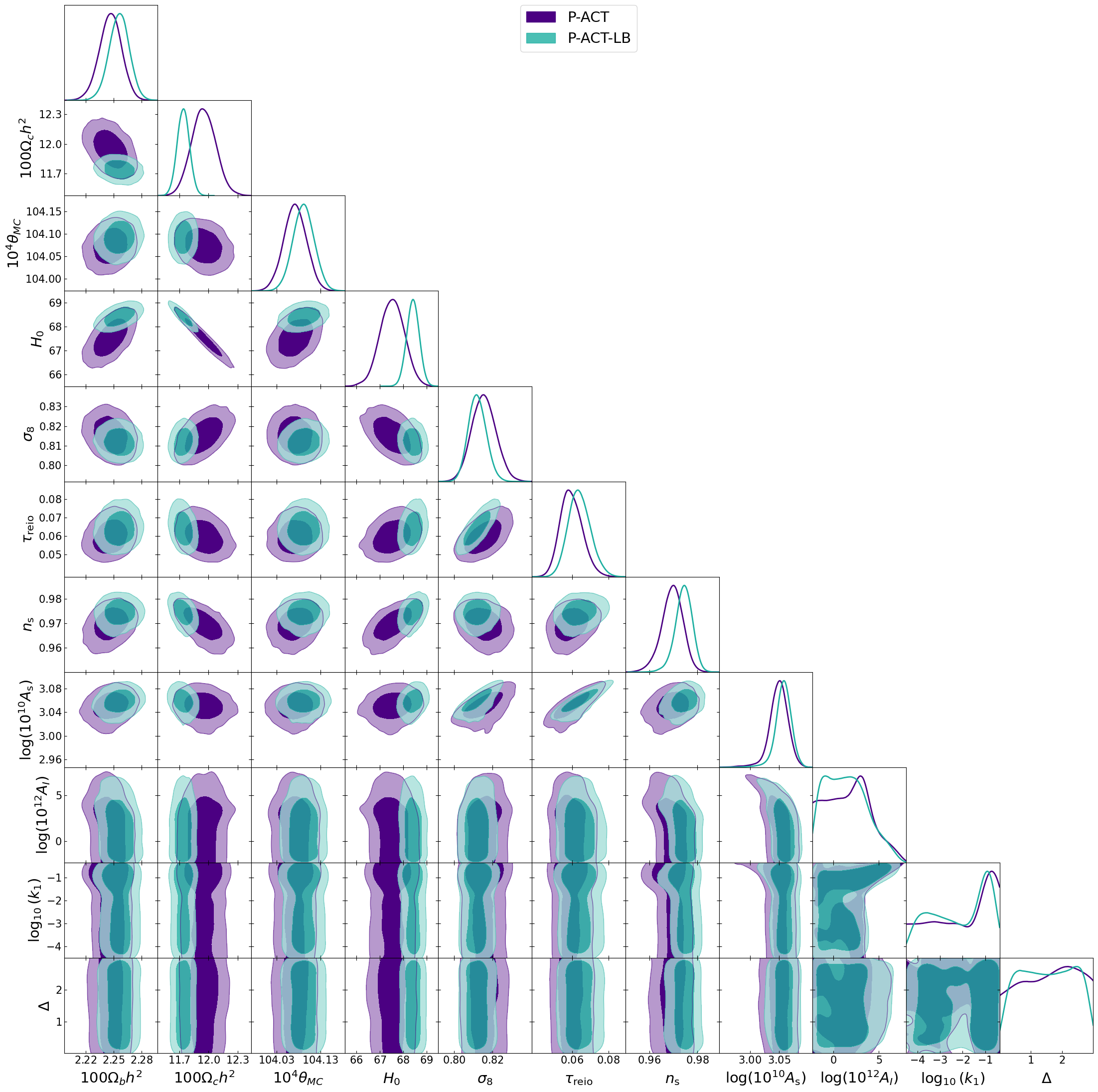}
    \caption{Marginalized 1D and 2D posterior distributions for the multiburst particle $P_{\mathcal{R}}(k)$ model for both P-ACT (indigo) and P-ACT-LB (teal). We show both the $\Lambda$CDM and particle burst parameters.}
    \label{fig: multi full}
\end{figure}

\begin{figure}
    \centering
    \includegraphics[width=0.8\linewidth]{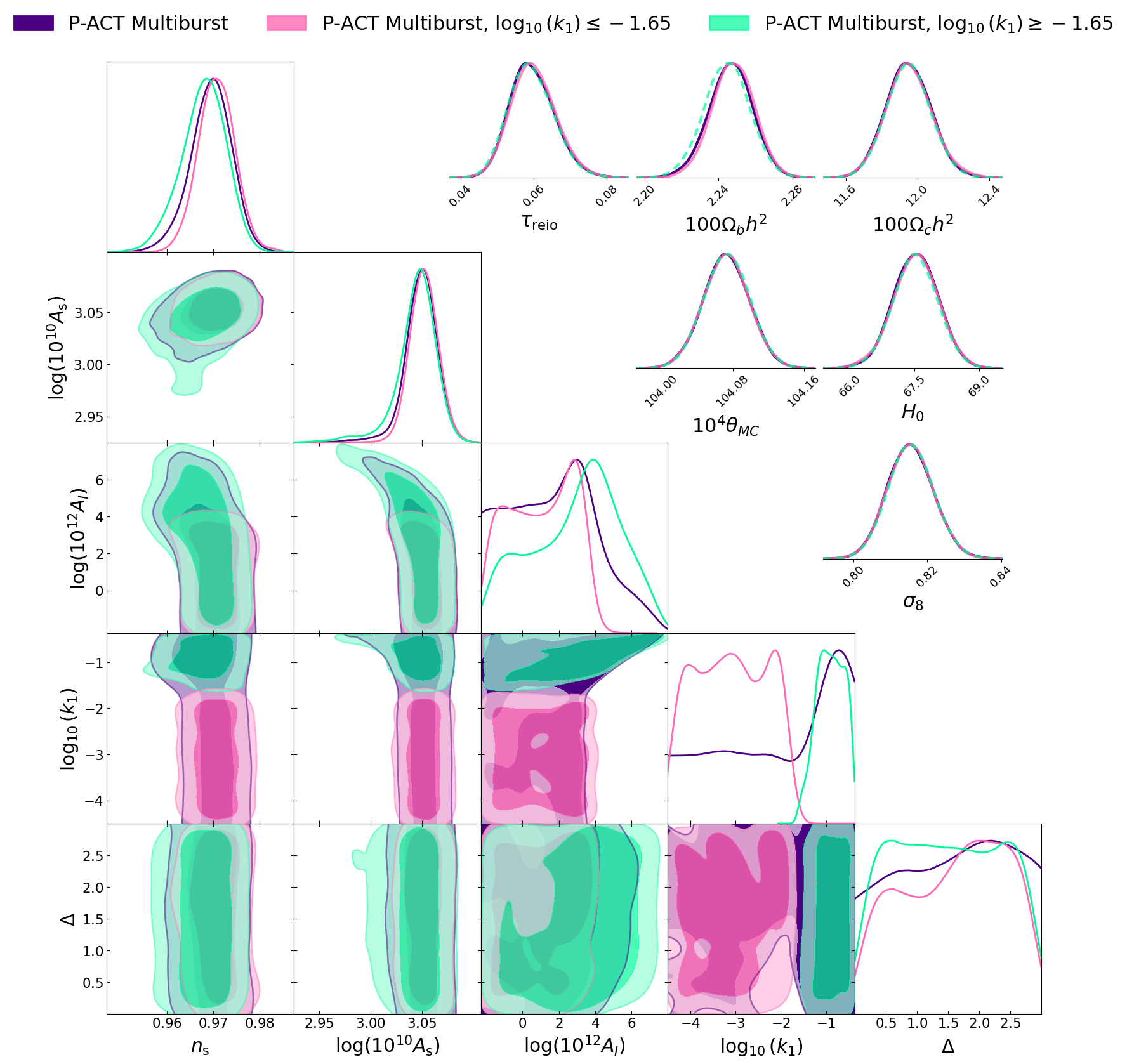}
    \caption{1D and 2D posterior distributions for the particle burst model for the different $k$-range priors. We show only the 1D posterior distributions for the parameters that do not have any significant changes from the base multiburst model (indigo). We show the results for the $k$-range priors of $\log_{10}(k_1) \leq -1.65$ (light green) and $\log_{10}(k_1) \geq -1.65$ (pink).}
    \label{fig: multi k range full}
\end{figure}

\begin{figure}
    \centering
    \includegraphics[width=0.8\linewidth]{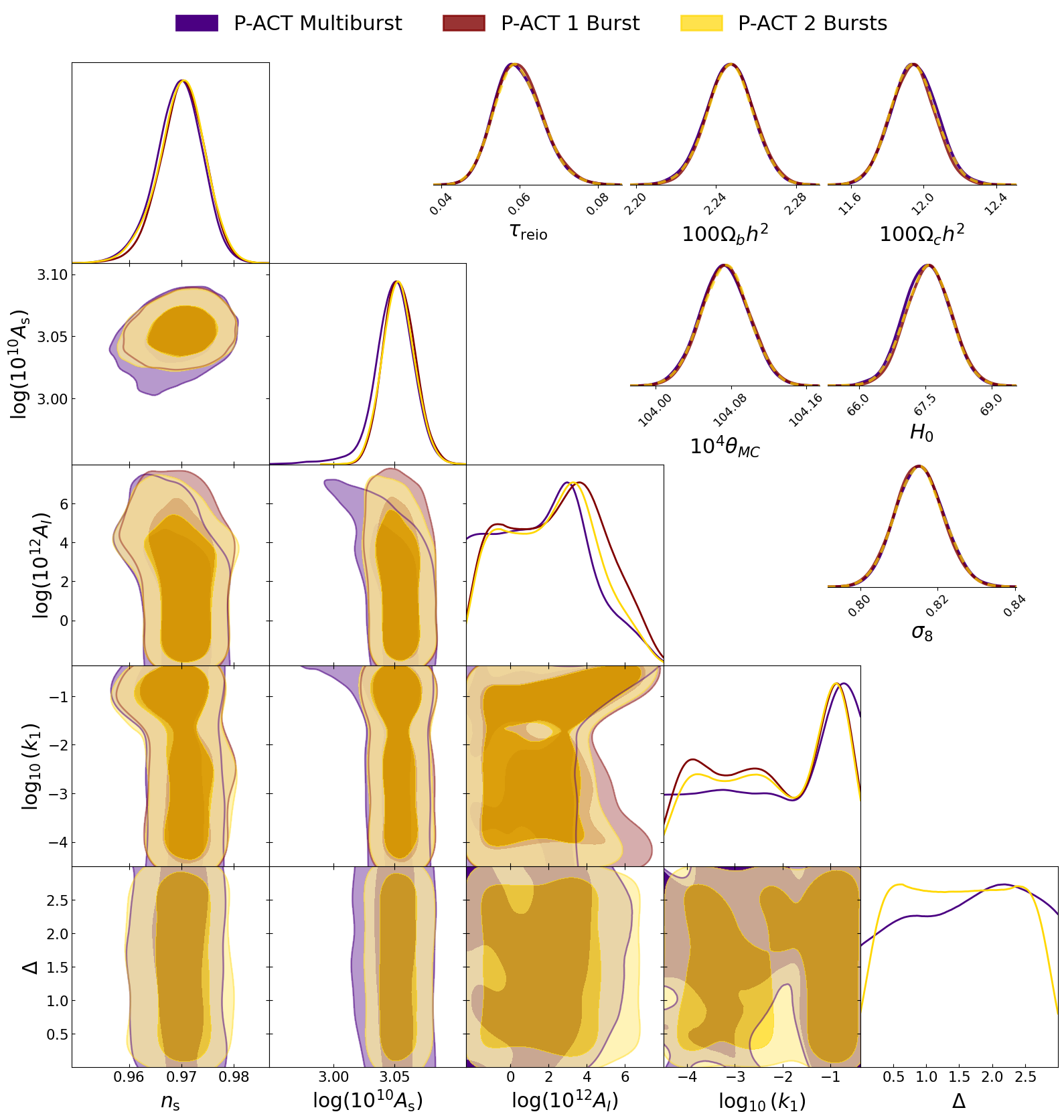}
    \caption{1D and 2D posterior distributions for the particle burst model for the different number of bursts. We show only the 1D posterior distributions for the parameters that do not have any significant changes from the base multiburst model (indigo). We show the results for 1 burst (maroon) and 2 bursts (yellow) of particle production.}
    \label{fig: multi 1b 2b full}
\end{figure}

Figure~\ref{fig: multi full} shows the 1D and 2D marginalized posterior distributions for all $\Lambda$CDM and particle burst parameters. We also provide the posterior distributions for the different $k$-range priors in Figure \ref{fig: multi k range full} and the number of bursts in Figure~\ref{fig: multi 1b 2b full}. For these $k$-range and number of burst plots, we show only the 1D posteriors for the parameters with minimal change. We again do not see a large shift in the $\Lambda$CDM parameters when comparing the $\Lambda$CDM-only runs, except for some small changes for $A_s$ and $n_s$ as expected. Additionally, we note that $\Delta$ is generally unconstrained in our analyses. 

\end{document}